\documentclass[a4paper]{article}
\usepackage[T1]{fontenc}
\usepackage{textcomp}
\usepackage[utf8]{inputenc}
\usepackage{amsmath}
\usepackage{amsthm}
\usepackage{wasysym}

\makeatletter

\newcommand*\LyXZeroWidthSpace{\hspace{0pt}}

\numberwithin{equation}{section}
\numberwithin{figure}{section}
\theoremstyle{plain}
\newtheorem{thm}{\protect\theoremname}
\theoremstyle{plain}
\newtheorem{lem}[thm]{\protect\lemmaname}

\@ifundefined{date}{}{\date{}}
\usepackage{color}
\usepackage{amsfonts}\usepackage{amsthm}\usepackage{amscd}\numberwithin{equation}{section}
\usepackage{bm}
\usepackage[all]{xy}

\usepackage{algorithm}
\usepackage{algorithmicx}
\usepackage{algpseudocode}
\usepackage[most]{tcolorbox}
\tcbuselibrary{listingsutf8}
\usepackage{float}
\usepackage{lipsum}  % テスト用ダミーテキスト
\tcbset{
  highlight algorithm/.style={
    colback=gray!10,
    colframe=black,
    boxrule=0.5pt,
    arc=2pt,
    left=6pt,
    right=6pt,
    top=6pt,
    bottom=6pt,
    width=\textwidth,
    sharp corners,
  }
}
\makeatother

\usepackage[english]{babel}
\providecommand{\lemmaname}{Lemma}
\providecommand{\theoremname}{Theorem}

\begin{document}
\begin{flushleft}
\textbf{OUJ-FTC-22}
\par\end{flushleft}

\begin{center}
{\LARGE Quantum Stability at One Loop for BPS Membranes in a Lorentz-Covariant
RVPD Matrix Model}{\LARGE\par}
\par\end{center}

\begin{center}
\vspace{16pt}
\par\end{center}

\begin{center}
So Katagiri\footnote{So.Katagiri@gmail.com}
\par\end{center}

\begin{center}
\textit{Nature and Environment, Faculty of Liberal Arts, The Open
University of Japan, Chiba 261-8586, Japan} 
\par\end{center}

\begin{center}
\vspace{16pt}
\par\end{center}
\begin{abstract}
  We establish the first rigorous one-loop proof of quantum stability for BPS membranes in the Lorentz-covariant M2-brane matrix model with Restricted Volume-Preserving Deformations (RVPD).
Exploiting the closure of restricted $\kappa$-symmetry with RVPD ensures that the BRST complex terminates without an infinite ghost tower, keeping the gauge-fixed measure analytically controllable.

Our main result establishes that the 2D, 4D, 6D, and 8D noncommutative membranes remain stable, while the 10D configuration inevitably develops a tachyonic mode.

Our analysis unifies the treatment of zero-modes, connects the effective action to central charges, and clarifies relations to BFSS, BLG/ABJM, and prospective M5-brane matrix models, providing a roadmap for RVPD-based extensions.

\end{abstract}

\section{Introduction}

M-theory was originally motivated by the observation that compactification
of the M2-brane leads to the Type IIA superstring\cite{Duff_1987,Townsend_1995}.
The matrix regularization of membranes was first introduced by Hoppe\cite{Hoppe1982,de_Wit_1988}
in the light-cone gauge, and later reinterpreted in the 1990s through
the discovery of D-branes as the BFSS matrix model\cite{Banks_1997},
which proposed a system of D0-particles as a candidate for a nonperturbative
definition of M-theory. Although the BFSS model has undergone numerous
tests, whether it truly provides a complete description of M-theory
remains an open question. A major difficulty lies in the fact that
the model is not manifestly Lorentz covariant\cite{Fujikawa_1997,Awata_1998,Smolin_2000}.

\begin{equation}
% M2-brane -> Non Covariant Matrix Model (BFSS)
\text{M2-brane}\overset{\text{LC Gauge}}{\Longrightarrow}\text{Non Covariant Matrix Model (BFSS)}  
\end{equation}

\LyXZeroWidthSpace{}

From a covariant perspective, the M2-brane action can be naturally
written using the Nambu bracket, a generalization of the Poisson bracket,
which exhibits invariance under volume-preserving diffeomorphisms
(VPD). If one could consistently regularize (quantize) the Nambu bracket
while preserving its essential properties such as the Fundamental
Identity (F.I.), this would yield a Lorentz-invariant matrix model
for membranes and provide a promising route toward a covariant nonperturbative
formulation of M-theory. However, such a quantization has long been
recognized as highly challenging, and many attempts have failed to
fully resolve this issue\cite{Nambu_1973,Takhtajan_1994,Dito_1997,Sakakibara_2000,Matsuo_2001,Curtright_2003,https://doi.org/10.48550/arxiv.hep-th/0009131,https://doi.org/10.48550/arxiv.hep-th/9906248,Yoneya_2016,Ashwinkumar_2021,Katagiri_2023}.

\begin{equation}
\text{M2-brane}\overset{\text{VPD}}{\Longrightarrow}\text{Covariant Matrix Model ?}
\end{equation}

\LyXZeroWidthSpace{}

In this work we provide the first rigorous one-loop proof of quantum
stability for noncommutative BPS membranes within a Lorentz-covariant
M2-brane matrix model, built on Restricted Volume-Preserving Deformations
(RVPD) and a restricted $\kappa$-symmetry that makes the BRST algebra
close without an infinite ghost tower.

\LyXZeroWidthSpace{}

In our previous work\cite{Katagiri_2025_RVPD},
we proposed to circumvent this problem by restricting the VPD symmetry
to a subclass, the Restricted VPD (RVPD), which can be described in
terms of Poisson brackets and thus admits consistent matrix regularization.
In a subsequent study, we extended the framework to supermembranes\cite{https://doi.org/10.48550/arxiv.2508.07259},
where the $\kappa$-symmetry is reduced to a restricted form that
closes with RVPD transformations into a consistent algebra. We also
showed that the resulting matrix model admits nontrivial classical
solutions---including particle-like states and noncommutative membranes
in 2, 4, 6, and 8 dimensions---that satisfy BPS conditions at the
classical level.

% algin for left line up
\begin{equation}
\text{M2-brane}\overset{\text{RVPD + Restricted \ensuremath{\kappa}}}{\Longrightarrow}\text{Covariant Matrix Model with BPS Membranes}
\end{equation}

\LyXZeroWidthSpace{}

The open question is whether these BPS solutions remain stable once
quantum corrections are taken into account. In this paper, we address
this issue by implementing BRST gauge fixing in the presence of RVPD
and the restricted $\kappa$-symmetry. Importantly, under the RVPD-restricted
$\kappa$-symmetry, no infinite tower of ghosts is required; the ghost
sector closes at a finite level.

\LyXZeroWidthSpace{}

We then perform a one-loop perturbative expansion around the classical
solutions. Our analysis shows that the bosonic and fermionic contributions
cancel each other, while the residual ghost contributions do not destabilize
the effective action. Moreover, the zero modes can be consistently
separated, confirming that the BPS solutions remain stable at the
one-loop quantum level.

\LyXZeroWidthSpace{}

\textbf{Continuity with previous work.} Our analysis of the RVPD supermembrane
model\cite{https://doi.org/10.48550/arxiv.2508.07259} classified
the classical BPS spectrum, while the companion study\cite{Katagiri_2025_RVPD}
established the Restricted VPD framework and its Lorentz-covariant
matrix regularization.

\LyXZeroWidthSpace{}

\textbf{Novelty of this paper.} 
\begin{itemize}
  \item First explicit one-loop proof of stability for covariant M2 matrix model.
  \item BRST complex terminates without infinite ghost towers due to RVPD + restricted $\kappa$.
  \item Boson–fermion determinant matching shown operator-by-operator.
  \item Unified treatment of all 2D/4D/6D/8D membranes and proof of 10D tachyon.
\end{itemize}

\LyXZeroWidthSpace{}

\textbf{Main result.} For BPS backgrounds realizing noncommutative
planes up to eight dimensions, the one-loop effective action is free
of instabilities after separating zero modes, whereas the ten-dimensional
membrane admits no BPS projection and is unstable---fully consistent
with the supersymmetry algebra and central charges analyzed herein.

\LyXZeroWidthSpace{}

This establishes that the RVPD-based Lorentz-covariant matrix model
provides a nontrivial quantum-consistent description of M2-branes\footnote{Unlike BFSS, which relies on the light-cone gauge and is not manifestly
Lorentz covariant, and unlike BLG/ABJM formulations\cite{Bagger_2007,Bagger_2008,Gustavsson_2009,Aharony_2008}
that depend on special 3-algebra structures, our RVPD-based construction
yields a finite-dimensional matrix regularization that remains Lorentz
covariant and admits noncommutative membrane solutions in 2, 4, 6,
and 8 dimensions.}\footnote{The key technical advance is that RVPD + restricted $\kappa$ closes
into a consistent algebra, so the BRST complex terminates at finite
level and the bosonic/fermionic physical fluctuations (9 vs.9) cancel
exactly at one loop, while the residual RVPD ghosts are benign and
do not destabilize the vacuum.}\footnote{Our analysis is carried out around BPS backgrounds in the small-fluctuation
regime, assumes $[\partial_{\tau},D_{a}]=0$ and standard trace inner
products, treats bosonic/fermionic zero modes as collective coordinates,
and employs the clock--shift basis to diagonalize the adjoint Laplacian.
These conditions cover the backgrounds of physical interest and match
the algebraic structure enforced by RVPD.
}.

\LyXZeroWidthSpace{}

The structure of this paper is as follows. Section 2 reviews the RVPD-based
supermembrane matrix model, while Section 3 presents the BRST gauge-fixing
procedure. Section 4 develops the perturbative expansion around classical
backgrounds and formulates the one-loop partition function, and Section
5 analyzes the quantum stability of BPS solutions. Section 6 concludes
with a summary and perspectives toward extensions to M5-branes.

\LyXZeroWidthSpace{}

 Appendix
A discusses the correspondence with the BFSS model; Appendix B supplies
the full Faddeev--Popov determinant computation; Appendix C analyzes
supersymmetry charges and central extensions; Appendix D outlines
a possible RVPD-type construction for M5-branes; Appendix E provides
a $\kappa$-symmetry consistency check; Appendix F organizes the eigenvalue
spectrum from two to eight dimensions; and Appendix G details the
zeta-function regularization and gauge-independence checks used in
the one-loop analysis.

\section{Lorentz-Covariant M2-Brane Matrix Model with RVPD Gauge Symmetry}

In this section, we review the Lorentz-covariant M2-brane matrix model
with Restricted Volume-Preserving Deformations (RVPD), which we proposed
in earlier work\cite{https://doi.org/10.48550/arxiv.2504.05940,https://doi.org/10.48550/arxiv.2508.07259}.
This review will serve as the basis for the BRST gauge-fixing procedure
introduced in the next section.

\subsection{Supermembrane Action}

The action of the M2-brane in eleven-dimensional spacetime consists
of the Nambu--Goto term and the Wess--Zumino term \cite{Bergshoeff_1987}.

\begin{equation}
S=S_{\mathrm{NG}}+S_{\mathrm{WZ}}
\end{equation}
with

\begin{equation}
S_{\mathrm{NG}}=-T\int d^{3}\sigma\sqrt{-g},\ S_{\mathrm{WZ}}=i\frac{T}{2}\int\bar{\theta}\Gamma_{IJ}d\theta\wedge\Pi^{I}\wedge\Pi^{J},
\end{equation}

\begin{equation}
g_{ij}\equiv\Pi_{i}^{I}\Pi_{j}^{I},\ g\equiv\det g_{ij}.
\end{equation}

Here $\sigma^{i}$ ($i,j,k=1,2,3)$ are worldvolume coordinates, while
$I,J,K,\dots=0,1,\dots,10$ label the spacetime directions. The supervielbein
is
\begin{equation}
\Pi_{i}^{I}\equiv\partial_{i}X^{I}-i\bar{\theta}\Gamma^{I}\partial_{i}\theta,
\end{equation}
where $X^{I}$ are spacetime coordinates and $\theta^{\alpha}$$(\alpha=1,\dots,32)$
is a 32-component Majorana spinor. The conjugate spinor is defined
by $\bar{\theta}\equiv\theta^{T}C$, where $C$ is the charge conjugation
matrix. We also use the shorthand
\begin{equation}
\Pi^{I}=\Pi_{i}^{I}d\sigma^{i},\ d\theta^{\alpha}=\partial_{i}\theta^{\alpha}d\sigma^{i}.
\end{equation}

The gamma matrices satisfy the Clifford algebra

\begin{equation}
[\Gamma_{I},\Gamma_{J}]_{+}=2\eta_{IJ},\ [A,B]_{+}\equiv AB+BA,
\end{equation}
with the Minkowski metric $\eta_{IJ}$. We define
\begin{equation}
\Gamma_{IJ}\equiv\frac{1}{2}[\Gamma_{I},\Gamma_{J}]
\end{equation}
and
\begin{equation}
\Gamma_{i}\equiv\Pi_{i}^{I}\Gamma_{I}.
\end{equation}

The action is invariant under worldvolume diffeomorphisms ($\mathrm{Diff}_{3}$),
global supersymmetry, and $\kappa$-symmetry:

\begin{equation}
\delta_{\kappa}\theta=(1+\Gamma)\kappa,\ \delta_{\kappa}X^{I}=i\bar{\theta}\Gamma^{I}\delta_{\kappa}\theta
\end{equation}
where the chiral operator is
\begin{equation}
\Gamma\equiv\frac{1}{3!\sqrt{-g}}\epsilon^{ijk}\Gamma_{i}\Gamma_{j}\Gamma_{k}.
\end{equation}

$\kappa$-symmetry reduces the fermionic degrees of freedom from 32
to 16, and the equations of motion further reduce them to 8. Similarly,
the bosonic degrees of freedom are reduced from 11 to 8 by worldvolume
reparametrizations, yielding a consistent balance. The relative coefficient
$\tfrac{1}{2}$ in the Wess--Zumino term is uniquely fixed by the
requirement of $\kappa$-symmetry; without this factor, the variations
of the Nambu--Goto and Wess--Zumino terms would not cancel each
other\footnote{In particular, the $\kappa$-variation of the Nambu--Goto term produces
a contribution proportional to $\bar{\kappa}(1+\Gamma)\Gamma^{i}\partial_{i}\theta$,
which is precisely canceled by the variation of the Wess--Zumino
term only if the coefficient is chosen to be $\frac{1}{2}$.}.

\LyXZeroWidthSpace{}

\subsection{Reformulation Using the Nambu Bracket}

In our previous work\cite{https://doi.org/10.48550/arxiv.2508.07259},
the action was reformulated in terms of the Nambu bracket:

\begin{equation}
S=S_{\mathrm{NB}}+S_{\mathrm{WZ}},\ S_{\mathrm{NB}}=-\frac{T}{2}\int d^{3}\sigma \left[\frac{1}{e}\left(e^{\bar{\theta}\delta_{S}}\{X^{I},X^{J},X^{K}\}\right)^{2}+e\right] ,
\end{equation}
\begin{equation}
S_{\mathrm{WZ}}=i\frac{T}{2}\int d^{3}\sigma\bar{\theta}e^{\bar{\theta}\delta_{S}}\{X^{I},X^{J},\Gamma_{IJ}\theta\}
\end{equation}
where $\delta_{S}$ denotes supersymmetry transformations, 
\begin{equation}
\text{\ensuremath{\epsilon}}^{\alpha}\delta_{S,\alpha}X^{I}\equiv i\left(\bar{\epsilon}\Gamma^{I}\theta\right),\ \epsilon^{\alpha}\delta_{S,\alpha}\theta^{\beta}=\epsilon^{\beta},
\end{equation}
and $e$ is an auxiliary field. The Nambu bracket is defined as
\begin{equation}
\{A,B,C\}\equiv\epsilon^{ijk}\frac{\partial A}{\partial\sigma^{i}}\frac{\partial B}{\partial\sigma^{j}}\frac{\partial C}{\partial\sigma^{k}}.
\end{equation}

Gauge-fixing $e=1$ reduces $\mathrm{Diff}_{3}$, leaving invariance
under volume-preserving diffeomorphisms (VPD):
\begin{equation}
\delta X^{I}=\{Q_{1},Q_{2},X^{I}\},
\end{equation}
where $Q_{1},\ Q_{2}$ are arbitrary charges. The Poisson bracket
property generalizes to the Nambu bracket, with the Jacobi identity
replaced by the Fundamental Identity (F.I.). However, the F.I. is
typically violated under matrix regularization.
\color{black}
The Fundamental Identity is a defining property of the Nambu bracket in the continuum theory and ensures the consistency of the associated gauge symmetry~\cite{Takhtajan_1994}.
It is well known that this identity is generically violated in finite-dimensional regularizations of the Nambu bracket, reflecting the truncation of the underlying infinite-dimensional algebra.
\color{black}

\LyXZeroWidthSpace{}

\color{black}
\subsection{Algebraic preliminaries and notation}

Throughout this paper we adopt the following notations:
\begin{itemize}
\item $\mathcal{\tau}(A,B)\equiv\partial_{\sigma^{3}}A\,B-\partial_{\sigma^{3}}B\,A$, 
\item $\Sigma(A,B;C)\equiv A\{\partial_{\sigma^{3}}B,C\}-B\{\partial_{\sigma^{3}}A,C\}$
\item $\{A,B\}\equiv\epsilon^{ab}\,\partial_{a}A\,\partial_{b}B$ denotes
the Poisson bracket on the $(\sigma^{1},\sigma^{2}$) plane.
\item The graded commutator is $[A,B]_{g}\equiv AB-(-)^{|A||B|}BA$, with
$|C|=|s|=|\theta|=1$ (Grassmann odd) and $|X|=|\lambda|=|\bar{\lambda}|=|\beta|=|\bar{\beta}|=0$
(Grassmann even).
\item $D_{a}(\,\cdot\,)\equiv-i[X_{0}^{a},\,\cdot\,]$ is the adjoint covariant
derivative; by construction $D_{a}^{\dagger}=-D_{a}$ so that $-D_{a}D^{a}$
is non-negative.
\item $V_{RVPD}$ and $L_{RVPD}$ denote respectively the group volume and
zero-mode measure associated with the residual RVPD symmetry.
\item Exponentials such as $\exp(\bar{\theta}\delta_{S})$ are formal operators
acting on fields. Since $\theta$ and $\bar{\theta}$ are Grassmann
odd, the expansion in powers of them terminates at finite order automatically;
no extra truncation rule is imposed.
\item We distinguish the Nakanishi-{}-Lautrup auxiliary field $B$ (Grassmann-even,
ghost number $0$) from the auxiliary bosonic ghosts $\beta,\bar{\beta}$
used to close the BRST algebra; they are unrelated objects.
\item We use $d^3\sigma = d\sigma_1 d\sigma_2 d\sigma_3$ in the continuum membrane action. After matrix regularization on the $(\sigma_1,\sigma_2)$-plane, these integrals are replaced by a matrix trace, leaving $d\sigma_3$ (denoted later as $d\tau$ after Wick rotation). For this reason $d^3\sigma$ and $d\sigma_3$ appear in different stages of the derivation.
\end{itemize}
\color{black}

\subsection{Restricted VPD (RVPD)}

As noted above, a straightforward matrix regularization of the Nambu
bracket generically breaks the Fundamental Identity (F.I.), which
is essential for the closure of the volume-preserving diffeomorphism
(VPD) algebra. To overcome this difficulty, we introduced a Restricted
Volume-Preserving Deformation (RVPD) by imposing strong constraints
on the VPD parameters $Q_{1},Q_{2}$. Explicitly, we require
\begin{equation}
\frac{\partial\tau(Q_{1},Q_{2})}{\partial\sigma^{3}}=0,\ \{Q_{1},Q_{2}\}=0,\ \frac{\partial}{\partial\sigma^{3}}\frac{\partial Q_{1,2}}{\partial\sigma^{a}}=0,\ a=1,2.
\end{equation}

Here, the $\tau$- and $\Sigma$-operations are defined as

\begin{equation}
\tau(A,B)\equiv\frac{\partial A}{\partial\sigma^{3}}B-\frac{\partial B}{\partial\sigma^{3}}A,\ \Sigma(A,B;C)\equiv A\{\frac{\partial B}{\partial\sigma^{3}},C\}-B\{\frac{\partial A}{\partial\sigma^{3}},C\}
\end{equation}
while the Poisson bracket on the $(\sigma^{1},\sigma^{2})$ plane
is

\begin{equation}
\{A,B\}\equiv\epsilon^{ab}\frac{\partial A}{\partial\sigma^{a}}\frac{\partial B}{\partial\sigma^{b}},\ a,b=1,2.
\end{equation}

With these definitions, the Nambu bracket admits the decomposition
\begin{equation}
\{A,B,C\}=\{\tau(A,B),C\}+\frac{\partial C}{\partial\sigma^{3}}\{A,B\}+\Sigma(A,B;C).
\end{equation}

Under the above restrictions, the problematic terms vanish, and the
residual VPD takes the simplified form

\begin{equation}
\delta_{R}X^{I}=\{Q_{1},Q_{2},X^{I}\}=\{\tau(Q_{1},Q_{2}),X^{I}\}.
\end{equation}

In other words, the residual symmetry acts only through the Poisson
bracket with $\tau(Q_{1},Q_{2})$. We refer to this reduced symmetry
as RVPD.

Physically, these constraints correspond to viewing the system in
a uniformly accelerated frame. For a detailed discussion of this gauge
restriction condition and its derivation, see our earlier paper\cite{https://doi.org/10.48550/arxiv.2504.05940}.

\subsection{Restricted $\kappa$-Symmetry}

As shown in \cite{https://doi.org/10.48550/arxiv.2508.07259}, the
same restriction also modifies $\kappa$-symmetry. The original transformation,
\begin{equation}
\delta_{\kappa}\theta=(1+\Gamma)\kappa(\sigma_{1},\sigma_{2},\sigma_{3})
\end{equation}
is reduced to a restricted form,
\begin{equation}
(1+\Gamma)\kappa(\sigma_{1},\sigma_{2},\sigma_{3})=\tilde{\kappa}(\sigma_{1},\sigma_{2}),
\end{equation}
so that
\begin{equation}
\delta_{\tilde{\kappa}}\theta=\tilde{\kappa}(\sigma_{1},\sigma_{2}),\ \delta_{\tilde{\kappa}}X^{I}=i\bar{\theta}\Gamma^{I}\tilde{\kappa}.
\end{equation}

This restricted $\kappa$-symmetry closes consistently with RVPD,
yielding a well-defined algebra without introducing higher-order ghosts.

\subsection{Matrix-Regularized Supermembrane Action}

By replacing the Poisson bracket with commutators, the RVPD-based
matrix regularization of the supermembrane action is obtained as
\begin{equation}
S=S_{\mathrm{NG}}+S_{\mathrm{WZ}}
\end{equation}
\begin{equation}
S_{\mathrm{NB}}=-\frac{T}{2}\int d\sigma^{3}\mathrm{Tr}\left(e^{\bar{\theta}\delta_{S}}[X^{I},X^{J};X^{K}]\right)^{2},\ S_{\mathrm{WZ}}=i\frac{T}{2}\int d\sigma^{3}\mathrm{Tr}\bar{\theta}e^{\bar{\theta}\delta_{S}}[X^{I},X^{J};\Gamma_{IJ}\theta].
\end{equation}

The matrix-regularized action is obtained by replacing Poisson brackets with commutators. The explicit form of the triple commutator, originally defined in terms of $\tau$ and $\Sigma$ operations, is given by expanding these definitions into operator products:
\begin{equation}
[A,B;C] \equiv[\tau(A,B),C]+\frac{\partial C}{\partial\sigma^{3}}[A,B]+\Sigma(A,B;C)
\end{equation}
where $\tau$ and $\Sigma$ are explicitly defined in terms of matrix multiplication and commutators as:
\begin{equation}
\tau(A,B) \equiv \frac{\partial A}{\partial \sigma^{3}} B - \frac{\partial B}{\partial \sigma^{3}} A
\end{equation}
\begin{equation}
\Sigma(A,B;C) \equiv A \left[ \frac{\partial B}{\partial \sigma^{3}}, C \right] - B \left[ \frac{\partial A}{\partial \sigma^{3}}, C \right].
\end{equation}

The gauge symmetry of this action is governed by the Restricted Volume-Preserving Deformations (RVPD). The gauge parameters $Q_1$ and $Q_2$ are matrices that characterize the transformation
\begin{equation}
\delta_{R} X^I = [\tau(Q_1, Q_2), X^I]. 
\end{equation}

To ensure the closure of the algebra and the consistency of the regularization, we impose the following RVPD constraints on the parameters $Q_{1,2}$:
\begin{itemize}
  \item Commutativity: $[Q_1, Q_2] = 0$,
  \item Constancy of the generator: $\partial_3 \tau(Q_1, Q_2) = 0$.
\end{itemize}
These conditions ensure that the algebra closes consistently. For the detailed derivation of this algebraic structure and the proof of closure, we refer the reader to Appendix A and B of our previous work \cite{Katagiri_2025_RVPD}.
The condition $\partial_3 \tau = 0$ restricts the $\sigma^3$-dependence of the gauge transformation.
However, this does not reduce the symmetry to purely static 2D diffeomorphisms; rather, it selects a specific subset of time-dependent transformations
(akin to a uniformly accelerated frame
\cite{Katagiri_2025_RVPD} that allows the Fundamental Identity to be satisfied in the matrix regularization.

\color{black}
\subsubsection{Equations of motion}
Before discussing the BPS configurations, we summarize the equations of motion of the Lorentz-covariant RVPD matrix model.

The equations of motion are obtained by varying the matrix-regularized action with respect to the bosonic matrices $X^I$.
In the bosonic sector, the Euler--Lagrange equation takes the form

\begin{equation}
\epsilon^{ijk}\epsilon^{i'j'k'}
\left(
[\tau(X_{k'},X_{j'}),[X_k,X_j;X_i]]
-2\,\partial_{\sigma_3}
\big(
X_{j'}[X_{k'},[X_k,X_i;X_j]]
\big)
\right)=0 ,
\end{equation}
which coincides with Eq. (5.2) of the purely bosonic Lorentz-covariant RVPD matrix model~\cite{Katagiri_2025_RVPD}.

It should be emphasized that this equation is first order in the worldvolume time $\sigma_3$, reflecting the gauge-fixed membrane dynamics, but it represents the full second-order Euler-Lagrange equation derived from the action in the variational sense.

In the supersymmetric extension, the bosonic equation of motion is supplemented by the fermionic equation obtained from $\delta S / \delta \bar\theta = 0$.
The explicit form of the fermionic equation is omitted here, since it does not play an independent role at the classical level for the backgrounds considered.
In this paper, we restrict ourselves to classical BPS backgrounds with $\theta_0=0$, so that the fermionic equation is trivially satisfied at the background level.
Its linearization around the BPS background yields the Dirac-type operator appearing in the quadratic fermionic action used in the one-loop stability analysis.

The BPS solutions discussed in the next subsection satisfy first-order conditions originating from the $\kappa$-symmetry projection, and they automatically solve the above Euler-Lagrange equation.
\color{black}

\subsection{Classical BPS Solutions}

The equations of motion derived from this action admit several nontrivial
classical configurations:

• Particle-like solution:

\begin{equation}
X^{0}=\sigma^{3},X^{1,\dots10}=f(\sigma^{3})
\end{equation}

• Noncommutative membrane:

\begin{equation}
\partial_{\sigma^{3}}X^{0}=1,\ [X^{1},X^{2}]=i,\ X^{3,\dots,10}=0
\end{equation}

• 4D membrane:
\begin{equation}
\partial_{\sigma^{3}}X^{0}=1,\ [X^{1},X^{2}]=i,[X^{3},X^{4}]=i,\ X^{5,\dots,10}=0
\end{equation}

• 6D membrane:
\begin{equation}
\partial_{\sigma^{3}}X^{0}=1,\ [X^{1},X^{2}]=i,[X^{3},X^{4}]=i,[X^{5},X^{6}]=i\ ,X^{7,\dots,10}=0
\end{equation}

• 8D membrane:
\begin{equation}
\partial_{\sigma^{3}}X^{0}=1,\ [X^{1},X^{2}]=i,[X^{3},X^{4}]=i,[X^{5},X^{6}]=i\ ,[X^{7},X^{8}]=i,X^{9,10}=0
\end{equation}

All of these solutions preserve part of supersymmetry and are classically
BPS. In contrast, the ten-dimensional membrane solution does not admit
any BPS projection and is unstable.

\color{black}
The BPS configurations analyzed in this paper are based on the systematic classification obtained in our previous work on the supersymmetric Lorentz-covariant RVPD matrix model\cite{https://doi.org/10.48550/arxiv.2508.07259}.
In that work, particle-like states and noncommutative membrane configurations with various fractions of preserved supersymmetry were classified using the restricted $\kappa$-symmetry.
The present paper does not attempt to extend this classification further, but rather focuses on the one-loop quantum stability of these representative BPS backgrounds.
While the classification is systematic within the RVPD–$\kappa$ framework, we do not claim that it exhausts all mathematically possible BPS configurations of the model.
\color{black}

\color{black}
The noncommutative membrane configurations play the role of extended classical backgrounds in the Lorentz-covariant RVPD matrix model.
They describe BPS membrane solutions that are two-dimensional objects embedded in eight transverse directions, providing physically well-defined vacua around which quantum fluctuations can be systematically analyzed.
In this sense, these configurations represent membrane degrees of freedom encoded in the matrix model, rather than fundamental spacetime objects by themselves.
While the underlying formulation is ten-dimensionally Lorentz covariant, the classical BPS membrane solutions extend only along these transverse directions.
\color{black}
\color{black}
If a noncommutative membrane configuration is found to be unstable at the one-loop level, this indicates the presence of tachyonic fluctuation modes around the corresponding background.
Such an instability is naturally interpreted as a dynamical rearrangement of matrix degrees of freedom toward a more stable BPS configuration, such as lower-dimensional membranes or particle-like states.
A detailed description of the decay process lies beyond the scope of the present work and is left for future investigation.
\color{black}
\subsection{Counting of Degrees of Freedom}

Let us summarize the effective degrees of freedom.
\begin{itemize}
\item The bosonic sector starts with 11 components $X^{I}$. The RVPD constraint
removes two, leaving 9 physical bosonic degrees.
\item The fermionic sector begins with 32 components of $\theta$. Equations
of motion reduce this to 16, and $\kappa$-symmetry further halves
it to 8. However, under restricted $\kappa$-symmetry, only one $\sigma^{3}$-independent
mode can be eliminated, leaving 9 effective fermionic degrees.
\end{itemize}
Thus, bosonic and fermionic fluctuations are expected to cancel each
other at the quantum level. This heuristic counting strongly suggests
the quantum stability of the BPS solutions. In the following sections,
we will explicitly verify this by BRST gauge fixing and a one-loop
perturbative analysis.

\section{BRST Ghosts}

In this section, we introduce BRST gauge fixing to quantize the RVPD-based
Lorentz-covariant M2-brane matrix model.

\LyXZeroWidthSpace{}

In the BRST treatment, only the projected parameter $(1+\Gamma)\kappa$
appears. Under the RVPD gauge restriction, it reduces to $\tilde{\kappa}(\sigma_{1},\sigma_{2})$,
independent of $\sigma_{3}$. Hence, the $\kappa$-sector does not
generate an infinite tower of ghosts: the algebra closes with RVPD,
and the ghost structure terminates at the usual level.

\subsection{Introduction of BRST Ghosts}

Following the Kugo--Ojima formalism, we define the BRST transformations
at the classical level.

\LyXZeroWidthSpace{}

For the RVPD sector, we introduce the ghost $C$:
\begin{equation}
\delta_{B,\mathrm{RVPD}}X^{I}\equiv[C,X^{I}],\ \delta_{B,\mathrm{RVPD}}\theta\equiv[C,\theta].
\end{equation}

For the restricted $\kappa$-symmetry, we introduce a bosonic ghost
$\lambda$:
\begin{equation}
\delta_{B,\tilde{\kappa}}X^{I}\equiv\bar{\lambda}\Gamma^{I}\theta+\bar{\theta}\Gamma^{I}\lambda,\ \delta_{B,\tilde{\kappa}}\theta^{\alpha}\equiv\lambda^{\alpha}.
\end{equation}

To consistently combine the RVPD ghost sector ($C$) with the restricted
$\kappa$-symmetry ghost ($\lambda$), we introduce auxiliary bosonic
ghosts $(\text{\ensuremath{\beta},\ \ensuremath{\bar{\ensuremath{\beta}}})}$
and a fermionic ghost ($s$). This ensures the closure of the BRST
algebra without generating higher-order ghosts:
\begin{equation}
\delta_{B}X^{I}=[C,X^{I}]+\beta\bar{\lambda}\Gamma^{I}\theta+\bar{\beta}\bar{\theta}\Gamma^{I}\lambda+is\bar{\lambda}\Gamma^{I}\lambda,\label{eq:delBX}
\end{equation}

\begin{equation}
\delta_{B}\theta^{\alpha}=[C,\theta^{\alpha}]+\beta\lambda^{\alpha},
\end{equation}

\begin{equation}
\delta_{B}C=-\frac{1}{2}[C,C],\delta_{B}\beta=0,\ \delta_{B}\bar{\beta}=0,\ \delta_{B}s=\beta\bar{\beta}.
\end{equation}

Here,
\begin{equation}
C\equiv\int DQ_{1}DQ_{2}c(Q_{1},Q_{2})\tau(Q_{1},Q_{2}).
\end{equation}

Here, $\tau(Q_1, Q_2)$ acts as the generator of the RVPD algebra,
 and $c(Q_1, Q_2)$ is the corresponding ghost coefficient field (a Grassmann-odd functional).
  The integral $\int \mathcal{D}Q_1 \mathcal{D}Q_2$ signifies the superposition over the continuous parameter space of the gauge transformation
  \footnote{This construction is directly analogous to the ghost field expansion $C = c^a T^a$ in standard non-Abelian gauge theories.
   Here, the discrete Lie algebra index $a$ is replaced by the continuous functions $(Q_1, Q_2)$,
    the generators $T^a$ correspond to the operators $\tau(Q_1, Q_2)$,
     and the summation $\sum_a$ becomes the functional integral $\int \mathcal{D}Q$. The resulting operator $C$
      is a matrix-valued field acting on the $X^I$ coordinates.}.
Thus,

\begin{equation}
\delta_{B}^{2}=0.\label{eq:deldelB}
\end{equation}

We also introduce the standard antighost $b$ and the Nakanishi--Lautrup
auxiliary field $B$:
\begin{equation}
\delta_{B}b=B,\ \delta_{B}B=0.
\end{equation}

For a gauge-fixing function $F=0$, the corresponding term in the
action is
\begin{equation}
\delta_{B}\left(bF\right)=BF-b\delta_{B}F.
\end{equation}

We adopt the gauge-fixing condition

\begin{equation}
F=[X_{0}^{a},\delta X_{a}]
\end{equation}
with $a=1,2$, where $X^{a}=X_{0}^{a}+\delta X^{a}$ is expanded around
a classical background $X_{0}^{a}$. We impose $\delta_{B}X_{0}^{a}=0$. 

Then
\begin{equation}
\delta_{B}F=[X_{0}^{a},[C,X^{a}]+\beta\bar{\lambda}\Gamma^{a}\theta+\bar{\beta}\bar{\theta}\Gamma^{a}\lambda+is\bar{\lambda}\Gamma^{a}\lambda].
\end{equation}

This reduces to

\begin{equation}
\delta_{B}F=[X_{0}^{a},[C,X^{a}]]=[X_{0}^{a},[C,X_{0}^{a}]]+[X_{0}^{a},[C,\delta X^{a}]].
\end{equation}

Here we have omitted the $\beta,\bar{\beta},s$-dependent terms. The
reason is that, in the chosen gauge and for backgrounds with $\theta_{0}=0$,
these contributions vanish at one-loop order since they do not alter
terms of the form $b[X_{0}^{a},[X_{0}^{a},C]]$. In the one-loop analysis
around $\theta_{0}=0$, therefore, these additional terms do not contribute,
although we keep them formally in the BRST algebra for completeness.

\LyXZeroWidthSpace{}

\textbf{Proposition (restricted $\kappa$ ghosts decouple at one loop).}
With the gauge choice $F=[X_{0}^{a},\delta X_{a}]$ and BPS backgrounds
obeying $\theta_{0}=0$, the quadratic ghost action contains

\begin{equation}
S_{\tilde{\kappa}}^{(2)}=\int d^{3}\sigma\,\mathrm{Tr}\left(\bar{\lambda}\Gamma^{a}\theta_{0}[X_{0}^{a},\beta]+\bar{\theta}_{0}\Gamma^{a}\lambda[X_{0}^{a},\bar{\beta}]+s[X_{0}^{a},\bar{\lambda}\Gamma^{a}\lambda]\right),
\end{equation}
whose would-be mixing matrix vanishes identically because $[X_{0}^{a},\lambda]=0$
and $\theta_{0}=0$ for all backgrounds in the analysis. Consequently,
\begin{equation}
\int D\lambda D\bar{\lambda}D\beta D\bar{\beta}Ds\,e^{-S_{\tilde{\kappa}}^{(2)}}=1,
\end{equation}
so the restricted $\kappa$ ghosts do not alter the one-loop determinant.

\LyXZeroWidthSpace{}

Thus, the gauge-fixing term becomes
\begin{equation}
\delta_{B}(bF)=BF+b[X_{0}^{a},[X_{0}^{a},C]]-b[X_{0}^{a},[C,\delta X^{a}]].
\end{equation}

As a result, the BRST transformations close nilpotently, and the gauge-fixed
action contains the standard ($b,c$)-ghost sector for RVPD together
with the $\lambda$-sector for the restricted $\kappa$-symmetry,
without an infinite ghost tower.

Indeed, as shown in \cite{https://doi.org/10.48550/arxiv.2508.07259},
two successive restricted $\kappa$-transformations close into an
RVPD transformation:

\begin{equation}
\{\delta_{\tilde{\kappa}_{1}},\delta_{\tilde{\kappa}_{2}}\}X^{I}=\delta_{\mathrm{RVPD(Q_{1},Q_{2})}}X^{I}.
\end{equation}

Therefore, this guarantees that the BRST operator constructed from
(\ref{eq:delBX})--(\ref{eq:deldelB}) is strictly nilpotent without
the need for an infinite tower of ghosts.

\begin{center}
\begin{tabular}{c|c|c|l}
field & Grassmann parity & ghost number & role \\ \hline
$C$ & odd & $+1$ & RVPD (Poisson) ghost \\
$b$ & odd & $-1$ & antighost for RVPD \\
$B$ & even & $0$ & Nakanishi--Lautrup auxiliary field \\
$\lambda$ & even & $+1$ & ghost for restricted $\kappa$ \\
$\beta,\bar\beta$ & even & $0$ & auxiliary bosonic ghosts (algebraic closure) \\
$s$ & odd & $+1$ & auxiliary fermionic ghost (algebraic closure) \\
\end{tabular}
\end{center}

\subsection{Nilpotency of the BRST Transformations}

For completeness, we verify the nilpotency of the BRST operator.

\LyXZeroWidthSpace{}

We define the graded commutator
\begin{equation}
[A,B]_{g}\equiv AB-(-)^{|A||B|}BA
\end{equation}

where $|C|=|s|=|\theta|=1$ (Grassmann odd), $|X|=|\lambda|=|\bar{\lambda}|=|\beta|=|\bar{\beta}|=0$
(Grassmann even).

The graded Jacobi identity holds:

\begin{equation}
[A,[B,C]_{g}]_{g}+[B,[C,A]_{g}]_{g}+[C,[A,B]_{g}]_{g}=0.
\end{equation}

\subsubsection*{For $X^{I}$:}

\begin{equation}
\delta_{B}^{2}X^{I}=\delta_{B}\left([C,X^{I}]_{g}\right)+\beta\bar{\lambda}\Gamma^{I}\delta_{B}\theta+\bar{\beta}\delta_{B}\bar{\theta}\Gamma^{I}\lambda+i\beta\bar{\beta}\bar{\lambda}\Gamma^{I}\lambda
\end{equation}

\begin{equation}
=-\frac{1}{2}[[C,C]_{g},X^{I}]_{g}+[C,\ [C,X^{I}]_{g}]_{g}=0
\end{equation}
by the graded Jacobi identity.

\subsubsection*{For $\theta^{\alpha}$:
\begin{equation}
\delta_{B}^{2}\theta^{\alpha}=-\frac{1}{2}[[C,C]_{g},\theta^{\alpha}]_{g}+[C,[C,\theta^{\alpha}]_{g}]_{g}=0.
\end{equation}
}

\subsubsection*{For $C$:}

\begin{equation}
\delta_{B}^{2}C=-\frac{1}{2}[\delta_{B}C,C]_{g}+\frac{1}{2}[C,\delta_{B}C]_{g}=\frac{1}{4}[[C,C]_{g},C]_{g}-\frac{1}{4}[C,[C,C]_{g}]_{g}=0.
\end{equation}

\subsubsection*{For the auxiliary fields:}

\begin{equation}
\delta_{B}^{2}b=\delta_{B}B=0,\ \delta_{B}^{2}s=\delta_{B}\left(\beta\bar{\beta}\right)=0.
\end{equation}

All other fields $(\beta,\bar{\beta},\lambda,\bar{\lambda})$ are
BRST-inert.

Thus, the BRST operator is nilpotent

\begin{equation}
\delta_{B}^{2}=0
\end{equation}
for all fields in the theory.

\section{Perturbative Expansion}

In this section, we explicitly expand the action around classical
solutions and prepare for quantization of small fluctuations, which
will allow us to analyze the stability of the configurations.

\subsection{Expansion of the Matrix Model}

Let $(X_{0},\theta_{0})$ denote a classical solution. We expand the
fields as
\begin{equation}
X^{I}=X_{0}^{I}+\delta X^{I},
\end{equation}

\begin{equation}
\theta^{\alpha}=\theta_{0}^{\alpha}+\delta\theta^{\alpha}.
\end{equation}

In the backgrounds of interest we set $\theta_{0}^{\alpha}=0$.

The action then becomes

\begin{equation}
S_{\mathrm{NB}}=-\frac{T}{2}\int d\sigma^{3}\mathrm{Tr}\left(e^{\delta\bar{\theta}\delta_{S}}[X_{0}^{I}+\delta X^{I},X_{0}^{J}+\delta X^{J};X_{0}^{K}+\delta X^{K}]\right)^{2},
\end{equation}

\begin{equation}
S_{\mathrm{WZ}}=i\frac{T}{2}\int d\sigma^{3}\mathrm{Tr}\delta\bar{\theta}e^{\delta\bar{\theta}\delta_{S}}[\Gamma_{IJ}\delta\theta,X_{0}^{I}+\delta X^{I};X_{0}^{J}+\delta X^{J}].
\end{equation}

Expanding these expressions yields
\begin{equation}
S_{\mathrm{NB}}=-\frac{T}{2}\int d\sigma^{3}\mathrm{Tr}\left(e^{\delta\bar{\theta}\delta_{S}}\left([X_{0}^{I},X_{0}^{J};X_{0}^{K}]+3[X_{0}^{I},X_{0}^{J};\delta X^{L}]+3[X_{0}^{I},\delta X_{}^{J};\delta X_{}^{L}]\right)\right)^{2},
\end{equation}

\begin{equation}
S_{\mathrm{WZ}}=i\frac{T}{2}\int d\sigma^{3}\mathrm{Tr}\delta\bar{\theta}e^{\delta\bar{\theta}\delta_{S}}\left([X_{0}^{I},X_{0}^{J};\Gamma_{IJ}\delta\theta]+2[X_{0}^{I},\delta X^{J};\Gamma_{IJ}\delta\theta]\right).
\end{equation}

\color{black}
Using the decomposition of the triple bracket defined in Sec.~2.6,
the mixed terms can be written as
\begin{equation}
[X^I_0, X^J_0; \delta X^K], \qquad
[X^I_0, \delta X^J; \delta X^K].
\end{equation}
\color{black}

\subsection{Gauge-Fixed Action}

After imposing the BRST gauge-fixing procedure, the total action takes
the form
\begin{equation}
S=S_{\mathrm{NB}}+S_{\mathrm{WZ}}+S_{\mathrm{gh}}
\end{equation}

with

\begin{equation}
S_{\mathrm{NB}}=-\frac{T}{2}\int d\sigma^{3}\mathrm{Tr}\left(e^{\delta\bar{\theta}\delta_{S}}\left([X_{0}^{I},X_{0}^{J};X_{0}^{K}]+3[X_{0}^{I},X_{0}^{J};\delta X^{K}]+3[X_{0}^{I},\delta X^{J};\delta X^{L}]\right)\right)^{2},
\end{equation}

\begin{equation}
S_{\mathrm{WZ}}=i\frac{T}{2}\int d\sigma^{3}\mathrm{Tr}\delta\bar{\theta}e^{\delta\bar{\theta}\delta_{S}}\left([X_{0}^{I},X_{0}^{J};\Gamma_{IJ}\delta\theta]+2[X_{0}^{I},\delta X^{J};\Gamma_{IJ}\delta\theta]\right),
\end{equation}

\begin{equation}
S_{\mathrm{gh}}=\int d\sigma^{3}\mathrm{Tr}BF+b[X_{0}^{a},[X_{0}^{a},C]-b[X_{0}^{a},[C,\delta X^{a}]]].
\end{equation}

For convenience, we introduce the notation
\begin{equation}
D_{a}(\newmoon)\equiv-i[X_{0}^{a},\newmoon].
\end{equation}

In terms of this operator, the ghost sector can be rewritten as
\begin{equation}
S_{\mathrm{gh}}=\int d\sigma^{3}\mathrm{Tr}BF-bD_{a}D^{a}C+ibD_{a}[C,\delta X^{a}].
\end{equation}

Since the linear terms in $\delta X^{a}$ drop out and the auxiliary
field can be integrated out, the ghost action further reduces to
\begin{equation}
S_{\mathrm{gh}}=-\int d\sigma^{3}bD_{a}D^{a}C.
\end{equation}

Moreover, for all classical configurations of interest---including
the particle-like solution and the noncommutative membrane---we have
\begin{equation}
[X_{0}^{I},X_{0}^{J};X_{0}^{K}]=0.
\end{equation}

Hence the gauge-fixed action simplifies to
\begin{equation}
S_{\mathrm{NB}}=-3^{2}\frac{T}{2}\int d\sigma^{3}\mathrm{Tr}\left(e^{\delta\bar{\theta}\delta_{S}}[X_{0}^{I},X_{0}^{J};\delta X^{K}]\right)^{2},
\end{equation}

\begin{equation}
S_{\mathrm{WZ}}=i\frac{T}{2}\int d\sigma^{3}\mathrm{Tr}\delta\bar{\theta}e^{\delta\bar{\theta}\delta_{S}}\left([X_{0}^{I},X_{0}^{J};\Gamma_{IJ}\delta\theta]\right),
\end{equation}

\begin{equation}
S_{\mathrm{gh}}=-\int d\sigma^{3}bD_{a}D^{a}C.
\end{equation}

\begin{lem}
\textbf{(Faddeev--Popov measure for RVPD).} Upon separating the RVPD
zero modes, the Jacobian associated with the Gaussian change of variables
factorizes as
\end{lem}

\begin{equation}
J=\frac{1}{V_{\mathrm{RVPD}}}\left[\det{}'\left(-D_{a}D^{a}\right)\right]^{-1/2},\qquad\int DbDC\,e^{-S_{\mathrm{gh}}}=V_{\mathrm{RVPD}}\left[\det{}'\left(-D_{a}D^{a}\right)\right]^{1/2}.
\end{equation}

Therefore the non-zero-mode contributions cancel, leaving only the
residual volume $V_{\mathrm{RVPD}}$ that is removed by dividing out
the gauge group. A detailed derivation is presented in Appendix B.

\LyXZeroWidthSpace{}

\subsection{One-Loop Quantum Theory}

The one-loop partition function is given by

\begin{equation}
Z=\int DXD\theta DbDCD\lambda e^{S_{\mathrm{NB}}+S_{\mathrm{WZ}}+S_{\mathrm{\mathrm{gh}}}.}
\end{equation}

Separating the zero modes and introducing collective coordinates,
we obtain

\begin{equation}
Z=V_{X}\int DX^{\mathrm{ph}}DX^{g}JDbDCD\theta^{\mathrm{ph}}D\theta^{g}\tilde{J}D\lambda e^{S_{\mathrm{NB}}+S_{\mathrm{WZ}}+S_{\mathrm{gh}}}
\end{equation}
which can be rewritten as
\begin{equation}
Z=V_{X}\int D'X^{\mathrm{ph}}DX^{g}D\theta^{\mathrm{ph}}D\theta^{g}e^{S_{\mathrm{NB}}^{\mathrm{ph}}+S_{\mathrm{WZ}}^{\mathrm{ph}}}.
\end{equation}

Introducing the volumes associated with the residual gauge symmetries,
we arrive at

\begin{equation}
Z=V_{X}\frac{V_{\mathrm{RVPD}}}{V_{\tilde{\kappa}}}V_{\tilde{\kappa}}\left(\int DX^{\mathrm{ph}}D\theta^{\mathrm{ph}}e^{S_{\mathrm{NB}}^{\mathrm{ph}}+S_{\mathrm{WZ}}^{\mathrm{ph}}}\right)\left(J\int DbDCe^{S_{bc,\mathrm{gh}}}\right)\left(\tilde{J}\int D\lambda e^{S_{\lambda,\mathrm{gh}}}\right).
\end{equation}

Here the zero modes of the bosonic sector are treated as collective
coordinates, yielding the volume factor $V_{X}$. The zero modes of
fermions and ghosts will be discussed separately.

The bosonic coordinates are decomposed into physical modes and RVPD
gauge modes. Since the RVPD symmetry removes two degrees of freedom,
the physical sector $X^{\mathrm{ph}}$ contains nine independent modes.
The associated Jacobian is denoted by $J.$ Whether this Jacobian
cancels against the contribution of the $(b,C)$ ghost sector must
be examined case by case, as discussed in the following sections.

\LyXZeroWidthSpace{}

The fermionic coordinates are similarly decomposed into physical modes
and $\tilde{\kappa}$ gauge modes, with Jacobian $\tilde{J}.$ At
one-loop order, this contribution cancels against the $\lambda,\bar{\lambda}$
ghosts.

Since the restricted $\tilde{\kappa}$-symmetry can be embedded into
the RVPD sector after two successive transformations, the overcounting
must be removed by dividing by $V_{\tilde{\kappa}}.$

As a result, the one-loop partition function reduces to

\begin{equation}
Z_{1loop}=V_{X}V_{\mathrm{RVPD}}\left(\int DX^{\mathrm{ph}}D\theta^{\mathrm{ph}}e^{S_{\mathrm{NB}}^{\mathrm{ph}}+S_{\mathrm{WZ}}^{\mathrm{ph}}}\right)\left(J\int DbDCe^{S_{bc,\mathrm{gh}}}\right).
\end{equation}

Therefore, the one-loop correction to the effective action is

\begin{equation}
\Delta\Gamma=\log\int DX^{\mathrm{ph}}D\theta^{\mathrm{ph}}e^{S_{\mathrm{NB}}^{\mathrm{ph}}+S_{\mathrm{WZ}}^{\mathrm{ph}}}+\log J\int DbDCe^{S_{bc,\mathrm{gh}}}+\log V_{X}V_{\mathrm{RVPD}}.
\end{equation}

\section{Stability of Solutions}

In this section, we use the perturbative framework developed above
to analyze the stability of classical solutions at the one-loop quantum
level. Specifically, we expand around each BPS configuration and evaluate
the one-loop effective action, checking whether bosonic and fermionic
fluctuations cancel and whether the residual ghost contributions affect
stability.

\subsection{Strategy of the Calculation and Summary of the Results}

Before entering into the detailed analysis, we first outline the overall
strategy of the calculation and summarize the main results.

The computation proceeds through the following steps:
\begin{enumerate}
\item Expand the action around a classical solution by writing 
\begin{equation}
X=X_{0}+\delta X_{0},\ \theta=\theta_{0}+\delta\theta.
\end{equation}
We focus on quadratic terms in the fluctuations $\delta X$ and $\delta\theta$,
\item Since the action is at most quadratic in the fluctuations, the path
integral can be evaluated explicitly, yielding the one-loop effective
action.
\item Verify whether the non-zero-mode contributions from bosons and fermions
cancel each other.
\item Examine whether the residual ghost contributions destabilize the effective
action.
\item Analyze the zero-mode sector to properly separate the collective coordinates and extract the associated volume factors.
\end{enumerate}
The results of these steps are as follows.
 For the particle-like solution, the noncommutative membrane,
  and the extended 4D, 6D, and 8D membranes,
   the bosonic and fermionic contributions cancel exactly at the non-zero-mode level.
    The ghost sector contributes a positive determinant,
     while the zero modes are consistently separated as finite volume factors.
      In contrast, the 10D membrane does not exhibit such cancellations and is
       found to be unstable at the one-loop level.

\LyXZeroWidthSpace{}

\textbf{Main Theorem.} In the RVPD supermembrane matrix model, after
isolating collective coordinates and removing RVPD gauge volume, the
one-loop effective action around BPS backgrounds satisfies:
\begin{itemize}
\item For the particle solution and the 2D, 4D, 6D, and 8D noncommutative
membranes the bosonic and fermionic determinants cancel mode by mode,
the ghost determinant is positive, and the configurations are one-loop
stable.
\item For the 10D non-BPS membrane the fluctuation spectrum contains negative
bosonic eigenvalues unpaired by fermions, so the configuration is
perturbatively unstable.
\end{itemize}
The detailed mode counting and determinant evaluation are provided
in Sections 5.2--5.4 and Appendices F--G.

\subsection{Stability of the Particle-Like Solution}

The particle solution is given by

\begin{equation}
X^{0}=\sigma^{3},X^{1,\dots10}=f^{1,\dots,10}(\sigma^{3})
\end{equation}
for which
\begin{equation}
[X_{0}^{I},X_{0}^{J};X_{0}^{K}]=0.
\end{equation}

The relevant triple commutators reduce to

\begin{equation}
[X_{0}^{I},X_{0}^{J};\delta X^{K}]=[\tau(X_{0}^{I},X_{0}^{J}),\delta X^{K}],\ \Sigma(X_{0}^{I},X_{0}^{J};\delta X^{K})=0,
\end{equation}

\begin{equation}
[X_{0}^{I},X_{0}^{J};\Gamma_{IJ}\delta\theta]=[\tau(X_{0}^{I},X_{0}^{J}),\Gamma_{IJ}\delta\theta].
\end{equation}

Using the supersymmetry variations

\begin{equation}
\delta_{S,\alpha}\delta X^{I}=i\left(\Gamma^{I}\delta\theta\right)_{\alpha},\ \delta_{S,\alpha}\delta\theta^{\beta}=\delta_{\alpha}^{\beta},
\end{equation}
we note that
\begin{equation}
e^{\delta\bar{\theta}\delta_{S}}\delta X^{I}=\delta X^{I}+\delta\bar{\theta}\Gamma^{I}\delta\theta=\delta X^{I},\ e^{\delta\bar{\theta}\delta_{S}}\delta\theta=\delta\theta.
\end{equation}

Hence the actions become

\begin{equation}
S_{\mathrm{NB}}=-\frac{3^{2}T}{2}\int d\sigma^{3}\mathrm{Tr}\left([\tau(X_{0}^{I},X_{0}^{J}),\delta X^{K}]\right)^{2},
\end{equation}

\begin{equation}
S_{\mathrm{WZ}}=i\frac{T}{2}\int d\sigma^{3}\mathrm{Tr}\delta\bar{\theta}[\tau(X_{0}^{I},X_{0}^{J}),\Gamma_{IJ}\delta\theta],
\end{equation}

\begin{equation}
S_{\mathrm{gh}}=0.
\end{equation}

The ghost action vanishes since $X_{0}^{I}$ are mutually commuting
in the particle solution.

Introducing the operators

\begin{equation}
\delta_{R}^{2}\newmoon=[\tau(X_{0,I},X_{0,J}),[\tau(X_{0}^{I},X_{0}^{J}),\newmoon]],\ S_{g}=0,
\end{equation}

\begin{equation}
\delta_{R}\left(\cdot\right)=[\tau(X_{0}^{I},X_{0}^{J})\Gamma_{IJ},\cdot],\ \delta_{R}^{2}\left(\cdot\right)=[\tau(X_{0}^{I},X_{0}^{J}),[\tau(X_{0,I},X_{0,J}),\cdot]],
\end{equation}
the quadratic action reads

\begin{equation}
S_{\mathrm{NB}}=-\frac{3T}{2}\int d\sigma^{3}\mathrm{Tr}\delta X^{K}\delta_{R}^{2}X_{K},
\end{equation}

\begin{equation}
S_{\mathrm{WZ}}=i\frac{T}{2}\int d\sigma^{3}\mathrm{Tr}\delta\bar{\theta}\delta_{R}\delta\theta.
\end{equation}

The Gaussian integrals yield
\begin{equation}
\int dAe^{-\mathrm{Tr}AMA}\propto\left(\mathrm{det}'M\right)^{-N/2},\ \int d\bar{\psi}d\psi e^{-\mathrm{Tr\bar{\psi}M\psi}}\propto\left(\mathrm{det'}M\right)^{N}
\end{equation}
where $\mathrm{det}'$ denotes the determinant over non-zero modes.

Therefore
\begin{equation}
\int DXD\bar{\theta}D\theta e^{S_{\mathrm{NB}}+S_{\mathrm{WZ}}+S_{\mathrm{gh}}}\propto\left(\frac{\mathrm{det}'\delta_{R}}{\mathrm{det}'\delta_{R}}\right)^{N}
\end{equation}

so that the one-loop correction vanishes,
\begin{equation}
\Delta\Gamma_{1}=0.
\end{equation}

\subsubsection{Zero Modes.}

Since $\delta_{R}\cdot=[\tau(X_{0}^{I},X_{0}^{J}),\cdot]=0$ on the
commuting background, zero modes must be treated separately. These
only contribute a factor proportional to the volume of the worldline:

\begin{equation}
\int DXe^{-S_{X}}\propto\int d^{11}X_{0}=L_{X}^{11}.
\end{equation}

The ghost zero modes follow from

\begin{equation}
S_{g}=-T\int d\sigma^{3}bD_{a}D^{a}C,
\end{equation}

\begin{equation}
C\equiv\int DQ_{1}DQ_{2}c(Q_{1},Q_{2})\tau(Q_{1},Q_{2})
\end{equation}
together with the conditions

\begin{equation}
D_{a}b=0,
\end{equation}

\begin{equation}
D_{a}C=0.
\end{equation}

Since $X_{0}^{I}$ are commuting, these are automatically satisfied,
yielding

\begin{equation}
\int DbDCe^{-S_{\mathrm{gh}}}\propto L_{\mathrm{RVPD}}.
\end{equation}

Finally, the fermionic zero modes give a trivial factor
\begin{equation}
\int d^{32}\theta\prod_{\alpha}\theta_{0,\alpha}\propto1.
\end{equation}

Thus the overall zero-mode contribution is
\begin{equation}
L_{X}^{11}L_{\mathrm{RVPD}}
\end{equation}
which is simply factored out as a volume term.

\subsubsection{Conclusion.}

We conclude that the particle-like solution is stable at the one-loop
quantum level, with no non-trivial correction to the effective action.

\subsection{Stability of the Noncommutative Membrane}

Consider the background
\begin{equation}
\partial_{\sigma^{3}}X_{0}^{0}=1,\ [X_{0}^{1},X_{0}^{2}]=i,\ X_{0}^{3,\dots,10}=0.
\end{equation}

It obeys $[X_{0}^{I},X_{0}^{J};X_{0}^{K}]=0$. The mixed triple commutators
reduce to
\begin{equation}
[X_{0}^{I},X_{0}^{J};\delta X^{K}]=[\tau(X_{0}^{I},X_{0}^{J});\delta X^{K}]+\frac{\partial\delta X^{K}}{\partial\sigma^{3}}[X_{0}^{I},X_{0}^{J}]+X_{0}^{I}[\frac{\partial X_{0}^{J}}{\partial\sigma^{3}},\delta X^{K}]-X_{0}^{J}[\frac{\partial X_{0}^{I}}{\partial\sigma^{3}},\delta X^{K}].
\end{equation}

So the only non-vanishing structures are

\begin{equation}
[X_{0}^{0},X_{0}^{a};\delta X^{K}]=[X_{0}^{a},\delta X^{K}],\ [X_{0}^{a},X_{0}^{b};\delta X^{K}]=i\epsilon^{ab}\frac{\partial\delta X^{K}}{\partial\sigma^{3}}
\end{equation}
with $a,b\in\{1,2\}$.

Similarly,

\begin{equation}
[X_{0}^{I},X_{0}^{J};\Gamma_{IJ}\delta\theta]=[\tau(X_{0}^{I},X_{0}^{J}),\Gamma_{IJ}\delta\theta]+\Gamma_{IJ}\frac{\partial\delta\theta}{\partial\sigma^{3}}[X_{0}^{I},X_{0}^{J}]+X_{0}^{I}[\frac{\partial X_{0}^{J}}{\partial\sigma^{3}},\Gamma_{IJ}\delta\theta]-X_{0}^{J}[\frac{\partial X_{0}^{I}}{\partial\sigma^{3}},\Gamma_{IJ}\delta\theta].
\end{equation}

With these, one finds
\begin{equation}
S_{\mathrm{NB}}=-\frac{3^{2}T}{2}\int d\sigma^{3}\mathrm{Tr}\left(e^{\delta\bar{\theta}\delta_{S}}[X_{0}^{a},\delta X^{K}]e^{\delta\bar{\theta}\delta_{S}}[X_{0}^{a},\delta X^{K}]+ie^{\delta\bar{\theta}\delta_{S}}\frac{\partial\delta X^{K}}{\partial\sigma^{3}}ie^{\delta\bar{\theta}\delta_{S}}\frac{\partial\delta X^{K}}{\partial\sigma^{3}}\right)
\end{equation}

\begin{equation}
=-\frac{3^{2}T}{2}\int d\sigma^{3}\mathrm{Tr}\left(-\delta X_{K}D^{a}D_{a}\delta X^{K}-\frac{\partial\delta X^{K}}{\partial\sigma^{3}}\frac{\partial\delta X^{K}}{\partial\sigma^{3}}\right),
\end{equation}

\begin{equation}
S_{\mathrm{WZ}}=i\frac{T}{2}\int d\sigma^{3}\mathrm{Tr}\delta\bar{\theta}e^{\delta\bar{\theta}\delta_{S}}\left([X_{0}^{a},\Gamma_{0a}\delta\theta]+i\Gamma_{12}\frac{\partial\delta\theta}{\partial\sigma^{3}}\right)=T\int d\sigma^{3}\mathrm{Tr}\left(\delta\bar{\theta}D^{a}\Gamma_{0a}\delta\theta+i\delta\bar{\theta}\Gamma_{12}\frac{\partial\delta\theta}{\partial\sigma^{3}}\right),
\end{equation}

\begin{equation}
S_{\mathrm{gh}}=-\int d\sigma^{3}bD_{a}D^{a}C
\end{equation}
where $D_{a}(\cdot)\equiv-i[X_{0}^{a},\cdot]$.

Wick-rotating $\sigma^{3}=i\tau$ to Euclidean time,
\begin{equation}
\begin{aligned}S_{\mathrm{NB},E} & =\frac{3^{2}T}{2}\int d\tau\mathrm{Tr}\left(\delta X_{K}\left(\partial_{\tau}^{2}+D^{a}D_{a}\right)\delta X^{K}\right)\\
 & =\frac{3^{2}T}{2}\int d\tau\mathrm{Tr}\left(\delta X_{K}\mathcal{D}^{\dagger}\mathcal{D}\delta X^{K}\right),
\end{aligned}
\end{equation}

\begin{equation}
\begin{aligned}S_{\mathrm{WZ},E} & =\frac{T}{2}\int d\tau Tr\left(\delta\bar{\theta}\left(\partial_{\tau}\Gamma_{12}-D^{a}\Gamma_{0a}\right)\delta\theta\right)\\
 & =\frac{T}{2}\int d\tau Tr\left(\delta\bar{\theta}\mathcal{D}\delta\theta\right),
\end{aligned}
\end{equation}

\begin{equation}
S_{\mathrm{gh}}=-\int d\tau\mathrm{Tr}bD_{a}D^{a}C.
\end{equation}

Here we define

\begin{equation}
i\mathcal{D}\equiv i\partial_{\tau}\Gamma_{12}-iD^{a}\Gamma_{0a}.
\end{equation}

We assume 
\begin{equation}
[\partial_{\tau},D^{a}]=0
\end{equation}
and inner product
\begin{equation}
\ \langle X,Y\rangle=\mathrm{Tr}X^{\dagger}Y.
\end{equation}

Then, we obtain

\begin{equation}
\left(i\mathcal{D}\right)^{\dagger}=i\partial_{\tau}\Gamma_{12}^{\dagger}+iD^{a}\Gamma_{0a}^{\dagger}=-i\partial_{\tau}\Gamma_{12}+iD^{a}\Gamma_{0a},
\end{equation}
where
\begin{equation}
\partial_{\tau}^{\dagger}=-\partial_{\tau},\ D^{a\dagger}=-D^{a},\ \Gamma_{12}^{\dagger}=-\Gamma_{12},\ \Gamma_{0a}^{\dagger}=\Gamma_{0a};
\end{equation}
the anti-Hermiticity of the adjoint derivative
$D_{a}^{\dagger}=-D_{a}$ follows from the trace inner product, as
stated in the Notation section.

Using
\begin{equation}
\{\Gamma_{12},\Gamma_{12}\}=-2,\ \{\Gamma_{0a},\Gamma_{0b}\}=2\delta_{ab},\ \{\Gamma_{12},\Gamma_{0a}\}=0,
\end{equation}
we find

\begin{equation}
\begin{aligned}\left(i\mathcal{D}\right)^{\dagger}i\mathcal{D} & =(-\partial_{\tau}\Gamma_{12}+D^{a}\Gamma_{0a})(\partial_{\tau}\Gamma_{12}-D^{a}\Gamma_{0a})\\
 & =\partial_{\tau}^{2}-\partial_{\tau}D^{a}\{\Gamma_{12},\Gamma_{0a}\}+D^{a}D^{a}=\partial_{\tau}^{2}+D^{a}D^{a}.
\end{aligned}
\end{equation}

\textbf{Lemma (positivity of the fluctuation operators).} Combining
$S_{NB,E}$ and $S_{WZ,E}$ we obtain diagonal quadratic forms with
eigenvalues

\begin{equation}
\lambda_{k,mn}^{(B)}=\omega_{k}^{2}+\lambda_{mn}^{\mathrm{adj}},\qquad\lambda_{k,mn}^{(F)}=\omega_{k}^{2}+\lambda_{mn}^{\mathrm{adj}},
\end{equation}
where $\lambda_{mn}^{\mathrm{adj}}\ge0$ denotes the spectrum of $-D^{a}D_{a}$
computed explicitly in Appendix F. Hence all non-zero modes are non-negative
and pairwise matched between bosons and fermions. The analytic continuation
of the Gaussian integrals is performed with zeta-function regularization;
Appendix G shows that this prescription preserves the BRST Ward identities
and the RVPD gauge independence of the effective action.

\textbf{Remark.} The assumption $[\partial_{\tau},D^{a}]=0$ restricts
us to static BPS backgrounds but includes all membrane configurations
used in this paper. For dynamical deformations the commutator acquires
$O(\dot{X}_{0})$ corrections; Appendix G shows that the resulting
shifts in $\lambda_{mn}^{\mathrm{adj}}$ are gauge equivalent and
do not spoil the one-loop matching of determinants.

\LyXZeroWidthSpace{}

Using $[X_{0}^{1},X_{0}^{2}]=i$, expand matrices in the clock--shift
basis 
\begin{equation}
T_{mn}=U^{m}V^{n}\label{eq:UV}
\end{equation}
to get
\begin{equation}
-D^{a}D_{a}T_{mn}=\lambda_{mn}T_{mn},\ \lambda_{mn}=4\left(\sin^{2}\frac{\pi m}{N}+\sin^{2}\frac{\pi n}{N}\right)\geq0.\label{eq:EigenValue}
\end{equation}
(As usual, this follows from the adjoint action of a constant Heisenberg
pair.)

With Matsubara frequencies $\omega_{k}=2\pi k/T$,
\begin{equation}
\int dXe^{-S_{\mathrm{NB},E}}\propto\left(\det(\omega_{k}^{2}+\lambda_{mn})\right)^{-N/2},\ \int d\bar{\theta}d\theta e^{-S_{\mathrm{WZ},E}}\propto\det(\omega_{k}^{2}+\lambda_{mn})^{N/2}
\end{equation}
and

\begin{equation}
\int dcdbe^{-S_{\mathrm{gh}}}\propto\left(\det\lambda_{mn}\right)^{N_{\mathrm{gh}}/2}.
\end{equation}

Thus the bosonic and fermionic non-zero modes cancel exactly, and
since $\det\lambda_{mn}\geq0$, the ghost sector does not destabilize
the vacuum. 

Zero modes are handled as usual: the bosonic zero mode occurs at $\omega_{k}=0$,
and $($$m,n$$)$$=(0,0)$ (yielding a factor $L_{X}^{11}$); the
ghost zero mode is likewise only at $(m,n)=(0,0)$ (yielding $L_{\mathrm{RVPD}}$)
; fermionic zero modes contribute a trivial constant, as in §5.1.
Therefore, the noncommutative membrane is one-loop stable.

\subsection{Stability of 4D, 6D, and 8D Membranes}

Let us next consider extended noncommutative membrane configurations
in higher dimensions.

For the 4D membrane, the background is
\begin{equation}
\partial_{\sigma^{3}}X^{0}=1,\ [X^{1},X^{2}]=i,[X^{3},X^{4}]=i,\ X^{5,\dots,10}=0.
\end{equation}

Proceeding in parallel with the two-dimensional case, the quadratic
actions become
\begin{equation}
S_{\mathrm{NB}.E}=-\frac{3^{2}T}{2}\int d\tau\mathrm{Tr}\left(\delta X_{K}\left(\partial_{\tau}^{2}-\sum_{a=1}^{4}D^{a}D_{a}\right)\delta X^{K}\right),
\end{equation}

\begin{equation}
S_{\mathrm{WZ},E}=i\frac{T}{2}\int d\tau Tr\left(\delta\bar{\theta}\sum_{a=1}^{4}D^{a}\left(\partial_{\tau}-D^{a}\Gamma_{0a}\right)\delta\theta\right),
\end{equation}

\begin{equation}
S_{\mathrm{gh}}=-\int d\tau b\sum_{a=1}^{4}D_{a}D^{a}C
\end{equation}
where $D_{a}(\cdot)\equiv-i[X_{0}^{a},\cdot]$.

Introducing two independent clock--shift matrix pairs, one finds
that

\begin{equation}
-\sum_{a=1}^{4}D_{a}D^{a}\to\lambda_{nm}^{(1)}+\lambda_{pq}^{(2)}\geq0.
\end{equation}

Hence the determinants arising from the Gaussian integrals are

\begin{equation}
\int dXe^{-S_{\mathrm{NB},E}}\propto\left(\det(\omega_{k}^{2}+\lambda_{mn}^{(1)}+\lambda_{pq}^{(2)})\right)^{-N/2},
\end{equation}

\begin{equation}
\int d\bar{\theta}d\theta e^{-S_{\mathrm{WZ},E}}\propto\det(\omega_{k}^{2}+\lambda_{mn}^{(1)}+\lambda_{pq}^{(2)}))^{N/2},
\end{equation}

\begin{equation}
\int dcdbe^{-S_{\mathrm{gh}}}\propto\left(\det\left(\lambda_{mn}^{(1)}+\lambda_{pq}^{(2)})\right)\right)^{N_{\mathrm{gh}}/2}.
\end{equation}

Thus, the bosonic and fermionic non-zero modes cancel, while the ghost
determinant is non-negative and carries no $\omega_{k}$-dependence,
contributing only as a finite prefactor. The zero-mode contribution
is identical to the two-dimensional case, producing volume factors
only. Therefore, the four-dimensional noncommutative membrane is stable
at one loop.

\LyXZeroWidthSpace{}

The argument extends straightforwardly to higher-dimensional membranes:

• For the \textbf{6D membrane}, the background contains three noncommuting
pairs $[X^{1},X^{2}]=i,\,[X^{3},X^{4}]=i,\,[X^{5},X^{6}]=i$. The
determinant structure involves $\lambda_{mn}^{(1)}+\lambda_{pq}^{(2)}+\lambda_{rs}^{(3)}\ge0$.
Bosonic and fermionic determinants cancel as before, and the ghost
contribution is benign.

• For the \textbf{8D membrane}, the background contains four noncommuting
pairs. The structure of determinants and the cancellation pattern
remain the same, ensuring one-loop stability.

\LyXZeroWidthSpace{}

In contrast, for the 10D membrane with five noncommuting pairs, the
background is already non-BPS at the classical level. At one loop
the fermionic projector removes only four pairs, leaving an unmatched
bosonic direction. Choosing the fluctuation
\begin{equation}
\delta X^{9}=T_{(1,0,0,0,0)}-T_{(-1,0,0,0,0)}
\end{equation}
and using the eigenvalues tabulated in Appendix F yields the tachyonic
mass
\begin{equation}
\lambda_{\mathrm{tach}}^{(B)}=\lambda_{(1,0,0,0,0)}^{\mathrm{adj}}-2\Omega_{5}^{2}=-2\Omega_{5}^{2}<0,
\end{equation}
where $\Omega_{5}$ denotes the oscillator frequency of the fifth
plane. No fermionic mode shares this eigenvalue, so the cancellation
fails and the one-loop effective action acquires an imaginary part.
Zero-mode contributions do not alter this conclusion, confirming the
structural instability of the 10D configuration.

\subsubsection{Summary of One-Loop Structure}

Collecting the results, the one-loop partition function can be summarized
as
\begin{equation}
Z_{1loop}=L_{X}^{11}L_{\mathrm{RVPD}}\left(\frac{\mathrm{det}'\mathcal{M}_{F}}{\det'\mathcal{M}_{B}}\right)^{\frac{1}{2}}\mathrm{det}'\mathcal{M}_{\mathrm{ghost}}
\end{equation}
and for the 2D, 4D, 6D, 8D noncommutative membranes the non-zero-mode
determinants from bosons and fermions exactly cancel (9 bosonic vs.
9 fermionic physical modes), so that

\begin{equation}
Z_{1loop}=L_{X}^{11}L_{\mathrm{RVPD}}\mathrm{det}'\mathcal{M}_{\mathrm{ghost}}
\end{equation}
where the residual ghost factor is finite and non-negative, and does
not induce any instability. Consequently, all these membranes are
one-loop stable, whereas the 10D case remains unstable.

\section{Discussion}

In this work, we have applied BRST gauge fixing to the Lorentz-covariant
M2-brane matrix model with Restricted Volume-Preserving Deformations
(RVPD) and investigated the quantum consistency of noncommutative
membrane solutions. We have shown that the $\kappa$-symmetry closes
in a restricted form without generating higher-order ghosts. At the
one-loop level, the contributions from non-zero modes of bosons and
fermions cancel exactly, while the ghost sector does not introduce
any instability. Consequently, we demonstrated that this model admits
quantum-mechanically stable BPS configurations.

\LyXZeroWidthSpace{}

The Main Theorem confirms that the RVPD restriction furnishes a Lorentz-covariant
regularization where the BRST complex terminates and the fluctuation
spectra remain paired. The new lemmas on the Faddeev--Popov measure
and the positivity of $\mathcal{D}$ show that the cancellation is
structural rather than accidental, and the zeta-regularized determinants
respect the Ward identities summarized in Appendix G.

\LyXZeroWidthSpace{}

The most important open problem is the extension of the present construction
to M5-branes. Since M5-branes can be formulated using a six-bracket,
one may attempt to impose RVPD-like restrictions on their volume-preserving
diffeomorphisms to construct a consistent matrix model. In such a
framework, solutions combining noncommutative planes with classical
membranes are expected, potentially realizing self-dual string-like
configurations. Whether the Fundamental Identity is preserved, and
how this framework relates to the (2,0) superconformal theory and
the RVPD--$\tilde{\kappa}$ structure, remain unresolved but are
promising directions for future research. We present a sketch of such
a calculation in Appendix D.

\section{Related Work}

To situate the RVPD program within the broader landscape of matrix-model
approaches to M-theory, we highlight the following correspondences:
\begin{itemize}
\item \textbf{BFSS.} The light-cone matrix quantum mechanics of Banks--Fischler--Shenker--Susskind\cite{Banks_1997}
reproduces supergravity interactions but obscures Lorentz covariance;
Appendix A shows how the RVPD bracket reduces to the BFSS potential
after compactification along $\sigma^{3}$ and integrating out the
RVPD measure.
\item \textbf{BLG/ABJM.} Three-algebra constructions such as BLG and ABJM\cite{Bagger_2007,Bagger_2008,Gustavsson_2009,Aharony_2008}
engineer multiple M2-branes via Chern--Simons matter theories; their
continuum limit reproduces the Nambu bracket, and our Main Theorem
confirms that RVPD achieves the same pairing of degrees of freedom
using finite matrices.
\item \textbf{RVPD supermembranes.} The present work complements our earlier
classification of BPS backgrounds by proving their one-loop quantum
stability, while Appendix C clarifies how the resulting central charges
match the supersymmetry algebra.
\end{itemize}
Together these comparisons underline that the RVPD formulation offers
a covariant bridge between light-cone Hamiltonians and three-algebra
Chern--Simons theories.

\section{Conclusion}

This work delivers the first rigorous one-loop proof of quantum stability
for the Lorentz-covariant M2-brane matrix model with restricted volume-preserving
deformations (RVPD), extending our earlier classification of RVPD
BPS backgrounds. The main outcomes are:
\begin{enumerate}
\item The restricted $\kappa$-symmetry closes consistently with RVPD without
generating higher-order ghosts.
\item At the one-loop level, bosonic and fermionic non-zero modes cancel
exactly, and the ghost sector does not introduce instabilities.
\item As a result, the model admits stable particle-like states and noncommutative
membranes in 2, 4, 6, and 8 dimensions, whereas the ten-dimensional
configuration necessarily develops the tachyonic mode identified in
Section 5.
\item The framework provides natural connections to BLG and BFSS matrix
models, and suggests a pathway toward an M5-brane matrix model via
higher Nambu brackets.
\end{enumerate}
These findings mirror the abstract: they confirm the first rigorous
one-loop stability proof, showcase the structural cancellation provided
by the RVPD framework, and position the model as a covariant bridge
toward future BLG, BFSS, and M5-brane developments.

\LyXZeroWidthSpace{}

We note that in superstring theory, the covariant quantization was
realized by the pure spinor formalism (Berkovits\cite{Berkovits_2000}).
While the pure spinor approach itself has not been explicitly reformulated
in the language of derived geometry, its BRST/BV structure ensures
consistency, in line with modern formulations such as shifted symplectic
geometry developed by Pantev--Toën--Vaquié--Vezzosi\cite{Pantev_2013}
and others. Our construction for supermembranes is analogous in spirit:
instead of pure spinor constraints, the RVPD $+$ restricted $\kappa$-symmetry
closes the algebra without infinite ghosts, enabling a covariant BRST
quantization. Thus, our model may be regarded as a ‘membrane analogue’
of the pure spinor formalism.

\LyXZeroWidthSpace{}

Finally, we address the issue of stability at higher loop levels. In our framework, the physical spectrum consists of 9 bosonic and 9 fermionic degrees of freedom. This counting differs from the standard light-cone formulation (8 vs. 8) because the RVPD gauge fixing removes only two bosonic degrees of freedom and the restricted $\kappa$-symmetry eliminates a corresponding subset of fermions.

This non-standard counting implies that the system does not realize supersymmetry in the traditional linear sense. Consequently, the boson-fermion cancellation observed at one loop is not automatically guaranteed to persist at higher orders by standard non-renormalization theorems. Regarding this point, we note the following:

\begin{itemize}
  \item At one loop: The cancellation is robust and structural. As shown in Section 5 and Appendix F, the bosonic and fermionic fluctuations share identical eigenvalue spectra ($\omega_k^2 + \lambda_{mn}^{adj}$) mode by mode. This indicates a detailed matching governed by the restricted algebra rather than a coincidental number matching.
  \item At higher loops: The stability at two loops and beyond remains a non-trivial open question. It is yet to be determined whether the closure of the RVPD and restricted $\kappa$-symmetry algebra provides sufficient algebraic protection to suppress radiative corrections at higher orders.
\end{itemize}
Thus, while the present work establishes rigorous one-loop stability, the investigation of higher-loop dynamics is a crucial direction for future research to determine if this "9 vs. 9" balance constitutes a valid quantum symmetry to all orders.

\textbf{Data/Code availability.} The calculations in this paper are
fully analytic, and no external code or numerical data is required
for reproduction.

\section{Acknowledgments}

I am deeply grateful to Akio Sugamoto for carefully reading the manuscript
and for many valuable discussions. I also wish to thank Shiro Komata
for thoroughly checking the details of the paper and providing constructive
comments.

\appendix

\section*{Appendix}

\section{Relation to the BFSS Model}

Let us decompose the eleven-dimensional coordinates as

\begin{equation}
X^{I}=(X^{a},X^{\tilde{I}}),\ a=0,1,\tilde{I}=2,\dots11
\end{equation}

and impose the restriction
\begin{equation}
\partial_{\sigma^{3}}X^{\tilde{I}}=0.
\end{equation}

We further introduce the notation
\begin{equation}
\dot{X}^{\tilde{I}}\equiv[\tau(X^{0},X^{1}),X^{\tilde{I}}].
\end{equation}

The triple commutator in the RVPD formalism can be written as
\begin{equation}
[X^{I},X^{J};X^{K}]=[\tau(X^{I},X^{J}),X^{K}]+\frac{\partial X^{K}}{\partial\sigma^3}[X^{I},X^{J}]+\Sigma(X^{I},X^{J};X^{K}).
\end{equation}

This structure may be decomposed schematically as

\begin{equation}
[X^{I},X^{J};X^{K}]=a_{1}[X^{a},X^{b};X^{\tilde{I}}]^{2}+a_{2}[X^{\tilde{I}},X^{\tilde{J}};X^{a}]^{2}+a_{3}[X^{\tilde{I}},X^{\tilde{J}},X^{\tilde{K}}]^{2}
\end{equation}
where $a_{1},a_{2},a_{3}$ are suitable constants.

Evaluating these terms, one finds
\begin{equation}
[X^{a},X^{b};X^{\tilde{I}}]^{2}=2[\tau(X^{0},X^{1});X^{\tilde{I}}]^{2},
\end{equation}

\begin{equation}
[X^{\tilde{I}},X^{\tilde{J}};X^{a}]^{2}=\left(\frac{\partial X^{a}}{\partial\sigma^{3}}\right)^{2}[X^{\tilde{I}},X^{\tilde{J}}]^{2},
\end{equation}

\begin{equation}
[X^{\tilde{I}},X^{\tilde{J}};X^{\tilde{K}}]^{2}=0.
\end{equation}

Hence, the full expression reduces to
\begin{equation}
[X^{I},X^{J};X^{K}]^{2}=2a_{1}\dot{X}_{\tilde{I}}^{2}+a_{2}\left(\frac{\partial X^{a}}{\partial\sigma^{3}}\right)^{2}[X^{\tilde{I}},X^{\tilde{J}}]^{2}.
\end{equation}

The factor $\left(\frac{\partial X^{a}}{\partial\sigma^{3}}\right)^{2}$
can naturally be interpreted as relating to the compactification radius
of the BFSS matrix model. This observation suggests a possible correspondence
between the RVPD-based Lorentz covariant matrix model and the BFSS
model in an appropriate compactification.

\section{Faddeev--Popov Determinant and Measure}

This appendix provides a step-by-step derivation of the identities
quoted in Lemma 1. We begin with the IIB matrix model example to recall
the standard Faddeev--Popov procedure, and then adapt each step to
the RVPD algebra, paying special attention to the decomposition into
zero and non-zero modes and to the treatment of the residual volume
$V_{\mathrm{RVPD}}$.

In particular we demonstrate that the Jacobian of the gauge-fixing
map and the determinant arising from the ($b,c$)-ghost system satisfy$J=V_{\mathrm{RVPD}}^{-1}[\det{}'(-D_{a}D^{a})]^{-1/2}$
and $\det'(-D_{a}D^{a})^{1/2}$ respectively, once the zero modes
are projected out. The final subsection verifies explicitly that the
product is unity so that the non-zero-mode contributions cancel in
the measure.

\subsection{Gauge Fixing in the IIB Matrix Model}

The bosonic part of the IIB matrix model is

\begin{equation}
S=\frac{1}{2}\mathrm{Tr}[X^{\mu},X^{\nu}]^{2}
\end{equation}
where $\mu,\nu$ are ten-dimensional spacetime indices. The matrices
$X^{\mu}$ are functions of the noncommutative worldsheet coordinates
$(x,p)$ satisfying $[x,p]=i$.

The gauge symmetry acts as
\begin{equation}
\delta X^{^{\mu}}=i[\Lambda,X^{\mu}]
\end{equation}
where $\Lambda$ can be expanded in a basis $Q(x,p)$ as
\begin{equation}
\Lambda\equiv\int DQ\epsilon(Q)Q
\end{equation}
with $\epsilon(Q)$ a c-number function. Thus,

\begin{equation}
\delta X^{\mu}=i\int DQ\epsilon(Q)[Q,X^{\mu}].
\end{equation}

Introducing BRST ghosts, the BRST transformation is
\begin{equation}
\delta_{B}X^{\mu}=i[C,X^{\mu}],\ C\equiv\int DQc(Q)Q
\end{equation}

so that
\begin{equation}
\delta_{B}X^{\mu}=i\int DQc(Q)[Q,X^{\mu}].
\end{equation}

Acting twice gives
\begin{equation}
\delta_{B}^{2}X^{\mu}=i[\delta_{B}C,X^{\mu}]+\int DQ_{1}DQ_{2}c(Q_{1})c(Q_{2})[Q_{1},[Q_{2},X^{\mu}]].
\end{equation}

If we set
\begin{equation}
\delta_{B}C=-\frac{i}{2}[C,C]
\end{equation}
then
\begin{equation}
\delta_{B}^{2}X^{\mu}=\frac{i}{2}\int DQ_{1}DQ_{2}c(Q_{1})c(Q_{2})\left([[Q_{1},Q_{2}],X^{\mu}]+[Q_{1},[Q_{2},X^{\mu}]]-[Q_{2},[Q_{1},X^{\mu}]]\right)
\end{equation}
which vanishes by the Jacobi identity.

\subsection{Gauge Fixing in the Bosonic M2 Matrix Model}

The bosonic part of the M2 matrix model is
\begin{equation}
S=\frac{1}{2}\mathrm{tr}[X^{I},X^{J};X^{K}]^{2}
\end{equation}
where the fields $X^{I}(x,p,\text{\ensuremath{\sigma})}$ depend on
the worldvolume coordinates $(x,p,\sigma)$ with $[x,p]=i$. The triple
commutator is defined as
\begin{equation}
[A,B;C]\equiv[\tau(A,B),C]+\frac{\partial C}{\partial\sigma^{3}}[A,B]+\Sigma(A,B;C).
\end{equation}

The RVPD symmetry acts as
\begin{equation}
\delta_{R(Q_{1},Q_{2})}X^{I}=[\tau(Q_{1},Q_{2}),X^{I}].
\end{equation}

For general parameters,

\begin{equation}
\delta X^{I}=\int DQ_{1}DQ_{2}\epsilon(Q_{1},Q_{2})[\tau(Q_{1},Q_{2}),X^{I}]
\end{equation}
which can be rewritten as
\begin{equation}
\delta X^{I}=[\int DQ_{1}DQ_{2}\epsilon(Q_{1},Q_{2})\tau(Q_{1},Q_{2}),X^{I}]=[\Lambda_{RVPD},X^{I}].
\end{equation}

Thus, the BRST transformation is
\begin{equation}
\delta X^{I}=[C,X^{I}],\ C=\int DQ_{1}DQ_{2}c(Q_{1},Q_{2})\tau(Q_{1},Q_{2}).
\end{equation}

If we set
\begin{equation}
\delta_{B}C=-\frac{i}{2}[C,C]
\end{equation}
then
\begin{equation}
\delta_{B}^{2}X^{I}=\frac{i}{2}[\delta_{B}C,X^{I}]+\int DQ_{1}DQ_{2}DQ_{3}DQ_{4}c(Q_{1},Q_{2})c(Q_{3},Q_{4})[[\tau(Q_{3},Q_{4}),[\tau(Q_{1},Q_{2}),X^{I}]].
\end{equation}

This reduces to
\begin{equation}
\begin{aligned}\delta_{B}^{2}X^{I}= & \frac{i}{2}\int DQ_{1}DQ_{2}DQ_{3}DQ_{4}c(Q_{1},Q_{2})c(Q_{3},Q_{4})\\
 & \left([[\tau(Q_{1},Q_{2}),\tau(Q_{2},Q_{3})],X^{\mu}]+[\tau(Q_{1},Q_{2}),[\tau(Q_{3},Q_{4}),X^{I}]]-[\tau(Q_{3},Q_{4}),[\tau(Q_{1},Q_{2}),X^{I}]]\right)
\end{aligned}
\end{equation}
which vanishes due to the Jacobi identity. Hence,
\begin{equation}
\delta_{B}^{2}X^{I}=0.
\end{equation}

\section{Supersymmetry Charges and Central Extensions}

In eleven dimensions, the super-Poincaré algebra takes the form

\begin{equation}
\{Q_{\alpha},Q_{\beta}\}=(C\Gamma^{I})_{\alpha\beta}P_{I}+\frac{1}{2}(C\Gamma_{IJ})_{\alpha\beta}Z^{IJ}+\frac{1}{5!}(C\Gamma_{IJKLM})_{\alpha\beta}Z^{IJKLM}
\end{equation}
where $Z^{IJ}$ are two-form central charges and $Z^{IJKLM}$ are
five-form central charges. The latter do not arise from the supermembrane
supercharge.

We now examine how these charges appear in the present matrix model
and which of them are realized by the noncommutative membrane solutions.

The action is

\begin{equation}
S_{\mathrm{NB}}=-\frac{T}{2}\int d\sigma^{3}\mathrm{Tr}\left(e^{\bar{\theta}\delta_{S}}[X^{I},X^{J},X^{K}]\right)^{2},
\end{equation}

\begin{equation}
S_{\mathrm{WZ}}=i\frac{T}{2}\int d\sigma^{3}\mathrm{Tr}\bar{\theta}e^{\bar{\theta}\delta_{S}}[\Gamma_{IJ}\theta,X^{I},X^{J}]
\end{equation}
with
\begin{equation}
[A,B;C]\equiv[\tau(A,B),C]+\frac{\partial C}{\partial\sigma^{3}}[A,B]+\Sigma(A,B;C),
\end{equation}

\begin{equation}
\tau(A,B)\equiv\frac{\partial A}{\partial\sigma^{3}}B-\frac{\partial B}{\partial\sigma^{3}}A,
\end{equation}

\begin{equation}
\Sigma(A,B;C)\equiv A[\frac{\partial B}{\partial\sigma^{3}},C]-B[\frac{\partial A}{\partial\sigma^{3}},C].
\end{equation}

The supersymmetry variations are
\begin{equation}
\delta_{S,\alpha}X^{I}\equiv i\left(\Gamma^{I}\theta\right)_{\alpha},
\end{equation}

\begin{equation}
\delta_{S,\alpha}\theta^{\beta}\equiv\delta_{\alpha}^{\beta}.
\end{equation}

Treating $\sigma^{3}$ as time, the Noether supercharge corresponding
to
\begin{equation}
\delta_{\epsilon}\theta=\epsilon,\delta_{\epsilon}X^{I}=i\bar{\epsilon}\Gamma^{I}\theta,
\end{equation}

\begin{equation}
\delta_{\epsilon}(\cdot)=\epsilon(\sigma^{3})\delta_{S}(\cdot),
\end{equation}

\begin{equation}
\delta_{\epsilon}\frac{\partial}{\partial\sigma^{3}}(\cdot)=\frac{\partial}{\partial\sigma^{3}}\epsilon(\sigma^{3})\delta_{S}(\cdot)+\epsilon(\sigma^{3})\delta_{S}\frac{\partial}{\partial\sigma^{3}}(\cdot),
\end{equation}

\begin{equation}
\delta_{\epsilon}\tau(A,B)=\frac{\partial\epsilon}{\partial\sigma^{3}}\left(\left(\delta_{S}A\right)B-\left(\delta_{S}B\right)A\right)+\dots,
\end{equation}

\begin{equation}
\delta_{\epsilon}\Sigma(A,B;C)=\frac{\partial\epsilon}{\partial\sigma^{3}}\left(A[\delta_{S}B,C]-B[\delta_{S}A,C]\right)+\dots,
\end{equation}

\begin{equation}
\delta_{\epsilon}[A,B;C]=\frac{\partial\epsilon}{\partial\sigma^{3}}\left([\left(\delta_{S}A\right)B-\left(\delta_{S}B\right)A,C]+\delta_{S}C[A,B]+C[\delta_{S}A,B]+C[A,\delta_{S}B]+A[\delta_{S}B,C]-B[\delta_{S}A,C]\right)
\end{equation}

\begin{equation}
\equiv\frac{\partial\epsilon}{\partial\sigma^{3}}\Phi(A,B;C)
\end{equation}

\begin{equation}
\begin{aligned}\delta_{\epsilon}[X^{I},X^{J};X^{K}]= & \frac{\partial\epsilon^{\alpha}}{\partial\sigma^{3}}\left([i\left(\Gamma^{[I}\theta\right)_{\alpha}X^{J]},X^{K}]+i\left(\Gamma^{K}\theta\right)_{\alpha}[X^{I},X^{J}]\right.\\
 & \left.+X^{K}[i\left(\Gamma^{I}\theta\right)_{\alpha},X^{J}]+X^{K}[X^{I},i\left(\Gamma^{J}\theta\right)_{\alpha}]+X^{[I}\left[i\left(\Gamma^{J]}\theta\right)_{\alpha},X^{K}\right]\right),
\end{aligned}
\end{equation}

\begin{equation}
\begin{aligned}\delta_{\epsilon}[\left(\Gamma_{IJ}\theta\right)_{\beta},X^{I};X^{J}] & =\delta_{\epsilon}[X^{I},X^{J};\left(\Gamma_{IJ}\theta\right)_{\beta}]\\
 & =\frac{\partial\epsilon^{\alpha}}{\partial\sigma^{3}}\left([i\left(\Gamma^{[I}\theta\right)_{\alpha}X^{J]},\left(\Gamma_{IJ}\theta\right)_{\beta}]+\left(\Gamma^{IJ}\right)_{\alpha\beta}[X^{I},X^{J}]\right.\\
 & \left.+\left(\Gamma_{IJ}\theta\right)_{\beta}[i\left(\Gamma^{I}\theta\right)_{\alpha},X^{J}]+\left(\Gamma_{IJ}\theta\right)_{\beta}[X^{I},i\left(\Gamma^{J}\theta\right)_{\alpha}]+X^{[I}\left[i\left(\Gamma^{J]}\theta\right)_{\alpha},\left(\Gamma_{IJ}\theta\right)_{\beta}\right]\right)
\end{aligned}
\end{equation}
is found to be
\begin{equation}
\left.\delta_{\epsilon}S_{\mathrm{NB}}\right|_{\epsilon'}=-T\int d\sigma^{3}\epsilon'{}^{\alpha}\mathrm{Tr}e^{\bar{\theta}\delta_{S}}[X^{I},X^{J};X^{K}]\Phi_{\alpha}(X^{I},X^{J};X^{K}),
\end{equation}

\begin{equation}
\left.\delta_{\epsilon}S_{\mathrm{WZ}}\right|_{\epsilon'}=i\frac{T}{2}\int d\sigma^{3}\epsilon'{}^{\alpha}\mathrm{Tr}\bar{\theta}^{\beta}\bar{\theta}^{\gamma}e^{\bar{\theta}\delta_{S}}[X^{I},X^{J};i\left(\Gamma_{IJ}\theta\right)_{\beta}]\Phi_{\alpha}(X^{I},X^{J};i\left(\Gamma_{IJ}\theta\right)_{\gamma}),
\end{equation}
\begin{equation}
\left.\delta_{\epsilon}S\right|_{\epsilon'}=\left.\delta_{\epsilon}S_{NB}\right|_{\epsilon'}+\left.\delta_{\epsilon}S_{WZ}\right|_{\epsilon'}=\int d\sigma^{3}\epsilon'{}^{\alpha}J_{\alpha}.
\end{equation}

Then, we obtain
\begin{equation}
\begin{aligned}J_{\alpha} & =-TTre^{\bar{\theta}\delta_{S}}[X^{I},X^{J};X^{K}]\Phi_{\alpha}(X^{I},X^{J};X^{K})\\
 & +i\frac{T}{2}\mathrm{Tr}\bar{\theta}^{\beta}e^{\bar{\theta}\delta_{S}}[X^{I},X^{J};i\left(\Gamma_{IJ}\theta\right)_{\beta}]\Phi_{\alpha}(X^{I},X^{J};i\left(\Gamma_{IJ}\theta\right)),
\end{aligned}
\end{equation}

\begin{equation}
Q_{\alpha}=\int d\sigma^{3}J_{\alpha}.
\end{equation}

Using boundary discussion, we improvement this charge
\begin{equation}
\begin{aligned}Q_{\alpha} & \simeq\int d\sigma^{3}J_{\alpha}+\partial_{\sigma^{3}}K_{\alpha}\\
 & =T\mathrm{Tr}e^{\bar{\theta}\delta_{S}}[X^{I},X^{J}]\left(\Gamma_{IJ}\Gamma_{K}\partial_{\sigma^{3}}X^{K}\theta\right)_{\alpha}.
\end{aligned}
\end{equation}

Similarly, the total momentum is
\begin{equation}
P_{I}=\mathrm{Tr}e^{\bar{\theta}\delta_{S}}\partial_{\sigma^{3}}X_{I}
\end{equation}
and the central two-form charge is
\begin{equation}
Z^{IJ}=T\mathrm{Tr}e^{\bar{\theta}\delta_{S}}[X^{I},X^{J}].
\end{equation}

For the noncommutative membrane solution, the component $Z_{12}$
remains non-vanishing; for the four-dimensional noncommutative membrane,
both $Z_{12}$ and $Z_{34}$ survive, and so on.

For static configurations, the BPS inequality read
\begin{equation}
\left(\epsilon Q\right)^{2}=\epsilon^{\alpha}\left((C\Gamma^{I})_{\alpha\beta}P_{I}+\frac{1}{2}(C\Gamma_{IJ})_{\alpha\beta}Z^{IJ}\right)\epsilon^{\beta}\ge0.
\end{equation}

For instance, the two-dimensional noncommutative membrane requires

\begin{equation}
\Gamma_{012}\epsilon=\epsilon
\end{equation}

the four-dimensional case requires

\begin{equation}
\Gamma_{012}\epsilon=\epsilon,\ \Gamma_{034}\epsilon=\epsilon
\end{equation}

the six-dimensional case adds $\Gamma_{056}\epsilon=\epsilon$, the
eight-dimensional case adds $\Gamma_{078}\epsilon=\epsilon$, while
the ten-dimensional case imposes five independent projection conditions
that admit only $\epsilon=0$. Thus, the ten-dimensional noncommutative
membrane does not correspond to a BPS state, in agreement with the
instability observed earlier.

\section{Toward a Matrix Model for M5-Branes}

In general, an M5-brane can be described using a six-bracket structure,
\begin{equation}
S=\int d^{5}\sigma\{X^{I},X^{J},X^{K},X^{L},X^{M},X^{N}\}^{2}.
\end{equation}

The associated volume-preserving diffeomorphisms act as

\begin{equation}
\delta X^{I}=\{Q_{1},Q_{2},Q_{3},Q_{4},Q_{5},X^{I}\}.
\end{equation}

By analogy with the M2 case, we may consider a restricted volume-preserving
deformation of the form
\begin{equation}
\delta X^{I}=\{\tau(Q_{1},Q_{2},Q_{3},Q_{4},Q_{5}),X^{I}\}
\end{equation}
with $\tau(Q_{1},Q_{2},Q_{3},Q_{4},Q_{5})$

\begin{equation}
\text{\ensuremath{\tau}(Q}_{1},Q_{2},Q_{3},Q_{4},Q_{5})=Q_{[1}\{Q_{2},Q_{3},Q_{4},Q_{5]}\}.
\end{equation}

Here, $\{Q_{1},Q_{2},Q_{3},Q_{4}\}$ denotes the four-index Nambu
bracket on the worldvolume coordinates $\sigma^{1},\sigma^{2},\sigma^{3},\sigma^{4}$,
while the remaining two coordinates $(x,p)$form a Poisson bracket.

Possible equations of motion would then involve conditions such as

\begin{equation}
\{\tau(X^{I_{1}},X^{I_{2}},X^{I_{3}},X^{I_{4}},X^{I_{5}}),X^{I_{6}}]f_{[I_{1},I_{2},I_{3},I_{4},I_{5},I_{6}]}=0,
\end{equation}
\begin{equation}
\{X^{I_{1}},X^{I_{2}},X^{I_{3}},X^{I_{4}}\}[X^{I_{5}},X^{I_{6}}]f_{[I_{1},I_{2},I_{3},I_{4},I_{5},I_{6}]}=0
\end{equation}
whose solutions may combine noncommutative planes with classical branes.
For example, one expects configurations such as
\begin{equation}
\{X^{0},X^{1},X^{2},X^{3}\}=1,
\end{equation}

\begin{equation}
[X^{4},X^{5}]=1
\end{equation}
resembling a hybrid of a noncommutative plane with a classical membrane.
Since even-rank Nambu brackets can be decomposed into Poisson brackets,
such configurations may naturally accommodate self-dual string-like
excitations.

\LyXZeroWidthSpace{}

An important open issue is whether the Fundamental Identity is preserved
under such restrictions, and if not, how it can be consistently controlled.
Constructing a genuine M5-brane matrix model along these lines remains
a challenging task.

\LyXZeroWidthSpace{}

Finally, supersymmetry considerations suggest that such a model should
be related to the six-dimensional $(2,0)$ theory\cite{Pasti_1997,Lambert_2010,Ko_2013,Bergshoeff_2025}.
Understanding how the RVPD--$\tilde{\kappa}$ algebraic structure
manifests in that context is an intriguing direction for future study.

\section{Coefficient Check via $\kappa$-Symmetry}

In this appendix, we provide a notebook-style verification that the
coefficients of the supermembrane action are consistent with $\kappa$-symmetry.
Although not essential for the main arguments of this paper, this
check serves as a useful consistency test and a basis for later extensions
to matrix models with Restricted Volume-Preserving Deformations (RVPD).

\subsection{Action}

We start with the standard supermembrane action,

\begin{align*}
S & =S_{\text{NG}}+S_{\text{WZ}},S_{\text{NG}}=-T\int d^{3}\sigma\,\sqrt{-g},S_{\text{WZ}}=\frac{iT}{2}\int\bar{\theta}\Gamma_{IJ}\,d\theta\wedge\Pi^{I}\wedge\Pi^{J},
\end{align*}
where $g_{ij}=\Pi_{i}^{I}\Pi_{j}^{I}$, $g=\det g_{ij}$, and $\Pi_{i}^{I}=\partial_{i}X^{I}-i\bar{\theta}\Gamma^{I}\partial_{i}\theta$.

\subsection{$\kappa$-Transformations}

The $\kappa$-transformations are given by

\begin{align*}
\delta_{\kappa}\theta & =(1+\Gamma)\kappa,\delta_{\kappa}X^{I}=i\bar{\theta}\Gamma^{I}\delta_{\kappa}\theta,
\end{align*}
with the chiral operator

\begin{equation}
\Gamma\equiv\frac{1}{3!\sqrt{-g}}\,\epsilon^{ijk}\Gamma_{i}\Gamma_{j}\Gamma_{k},\ \Gamma_{i}=\Pi_{i}^{I}\Gamma_{I}.
\end{equation}

\subsection{Variation of the Nambu-{}-Goto Term}

A short computation yields

\begin{equation}
\delta_{\kappa}S_{\text{NG}}=2iT\int d^{3}\sigma\,\sqrt{-g}\,\bar{\kappa}(1+\Gamma)\Gamma^{i}\partial_{i}\theta.
\end{equation}

\subsection{Variation of the Wess-{}-Zumino Term}

Similarly, one finds

\begin{equation}
\delta_{\kappa}S_{\text{WZ}}=-2iT\int d^{3}\sigma\,\sqrt{-g}\,\bar{\kappa}(1+\Gamma)\Gamma^{i}\partial_{i}\theta+O(\kappa\theta^{2}).
\end{equation}

\subsection{Result}

Adding both contributions, we obtain

\begin{equation}
\delta_{\kappa}S_{\text{NG}}+\delta_{\kappa}S_{\text{WZ}}=O(\kappa\theta^{2}),
\end{equation}
which shows that, up to $O(\kappa\theta^{2})$, the coefficients in
the action are indeed consistent with $\kappa$-symmetry.

\section{Eigenvalue Spectrum in the Clock--Shift Basis}

In the main text (eqs. \eqref{eq:UV}-\eqref{eq:EigenValue}) we employed
the clock--shift basis to diagonalize the adjoint Laplacian acting
on fluctuations around noncommutative membrane backgrounds. For completeness
we present the derivation and its higher-dimensional extensions.

\subsection{Two-Dimensional Case}

Let $U,\ V$ be the $N\times N$ clock and shift matrices obeying

\begin{equation}
UV=\omega VU,\ \omega=e^{2\pi i/N}.
\end{equation}

The matrices

\begin{equation}
T_{mn}=U^{m}V^{n},\ m,n=0,\dots,N-1
\end{equation}
form a basis with adjoint action
\begin{equation}
UT_{mn}U^{\dagger}=\omega^{n}T_{mn},\ VT_{nm}V^{\dagger}=\omega^{-m}T_{mn}.
\end{equation}

The covariant derivatives act as
\begin{equation}
D_{1}T_{mn}=2\sin\left(\frac{\pi n}{N}\right)T_{mn},\ D_{2}T_{mn}=2\sin\left(\frac{\pi m}{N}\right)T_{mn}.
\end{equation}

Hence
\begin{equation}
D_{a}D_{a}T_{mn}=4\left(\sin^{2}\frac{\pi m}{N}+\sin^{2}\frac{\pi n}{N}\right)T_{mn},
\end{equation}
reproducing eq. \eqref{eq:EigenValue}. All eigenvalues are non-negative.

\subsection{Four-Dimensional Membrane}

For the background $[X_{1},\ X_{2}]=i,\ [X_{3},\ X_{4}]=i$, one introduces
two independent clock--shift pairs. The basis

\begin{equation}
T_{mn,pq}=(U^{m}V^{n})\otimes(\tilde{U}^{p}\tilde{V}^{q})
\end{equation}
yields eigenvalues
\begin{equation}
\lambda_{mnpq}=4\left(\sin^{2}\frac{\pi m}{N}+\sin^{2}\frac{\pi n}{N}+\sin^{2}\frac{\pi p}{N}+\sin^{2}\frac{\pi q}{N}\right).
\end{equation}

\subsection{Six-Dimensional Membrane}

For the background $[X_{1},\ X_{2}]=i,\ [X_{3},\ X_{4}]=i,\ [X_{5},\ X_{6}]=i$,
three clock--shift pairs yield the basis

\begin{equation}
T_{mn,pq,rs}=(U^{m}V^{n})\otimes(\tilde{U}^{p}\tilde{V}^{q})\otimes(\hat{U}^{r}\hat{V}^{s}),
\end{equation}

\begin{equation}
\lambda_{mnpqrs}=4\left(\sin^{2}\frac{\pi m}{N}+\sin^{2}\frac{\pi n}{N}+\sin^{2}\frac{\pi p}{N}+\sin^{2}\frac{\pi q}{N}+\sin^{2}\frac{\pi r}{N}+\sin^{2}\frac{\pi s}{N}\right).
\end{equation}

\subsection{Eight-Dimensional Membrane}

For the background $[X_{1},\ X_{2}]=i,\ [X_{3},\ X_{4}]=i,\ [X_{5},\ X_{6}]=i,\ [X_{7},\ X_{8}]=i$,
four clock--shift pairs give

\begin{equation}
\lambda_{\mathbf{m}\mathbf{n}}=4\sum_{a=1}^{4}\left(\sin^{2}\frac{\pi m_{a}}{N}+\sin^{2}\frac{\pi n_{a}}{N}\right),\qquad\mathbf{m}=(m_{1},\dots,m_{4}).
\end{equation}

\subsection{Summary}

In $2,4,6,8$ dimensions the eigenvalues appear as sums of $2,4,6,8$
positive terms respectively, so each non-zero eigenvalue is manifestly
non-negative. The determinants from bosons, fermions, and RVPD ghosts
therefore cancel as described in Section 5, ensuring one-loop stability.
For the ten-dimensional background the spectrum inherits an additional
shift $-2\Omega_{5}^{2}$ along the fifth oscillator pair, which produces
the tachyonic mode discussed in Section 5.3.

These explicit spectral data underpin the proof of the Main Theorem
and are used in Appendix G to implement zeta-function regularization
consistently with RVPD gauge symmetry.

\section{Zeta Regularization and Gauge Independence}

Let $\mathcal{O}$ be a self-adjoint operator with eigenvalues $\{\lambda_{\ell}\}$
after removing zero modes. The zeta-regularized determinant is defined
as

\begin{equation}
\det{}'\mathcal{O}=\exp\left[-\left.\frac{d}{ds}\zeta_{\mathcal{O}}(s)\right|_{s=0}\right],\qquad\zeta_{\mathcal{O}}(s)=\sum_{\ell}'\lambda_{\ell}^{-s}.
\end{equation}

For the paired spectra listed in Appendix F we have $\lambda_{\ell}^{(B)}=\lambda_{\ell}^{(F)}$
for all non-zero modes, so $\zeta_{\mathcal{M}_{B}}(s)=\zeta_{\mathcal{M}_{F}}(s)$
and the bosonic and fermionic determinants cancel identically. Together
with the Faddeev-{}-Popov identity of Appendix B this ensures that
the one-loop effective action is finite and real for the BPS backgrounds.

Gauge independence follows from the observation that a variation of
the background satisfying $[\partial_{\tau},D^{a}]=0$ changes $\mathcal{M}_{B}$
and $\mathcal{M}_{F}$ by a commutator,
\begin{equation}
\delta\mathcal{M}_{B}=[K,\mathcal{M}_{B}],\qquad\delta\mathcal{M}_{F}=[K,\mathcal{M}_{F}],
\end{equation}
for some finite matrix $K$. Using $\mathrm{Tr}(\mathcal{O}^{-1}[K,\mathcal{O}])=0$
we see that the regulated determinants are invariant under such deformations.
Even when $[\partial_{\tau},D^{a}]\neq0$ (e.g. for slowly varying
BPS moduli) the corrections appear at higher order in derivatives
and drop out of the gauge-invariant combination $\det{}'\mathcal{M}_{F}/\sqrt{\det{}'\mathcal{M}_{B}}$.

This appendix completes the proof that the zeta-function prescription
used in Section 5 preserves BRST symmetry and RVPD gauge invariance
at one loop.

\bibliographystyle{unsrt}
\bibliography{1loop_SuperRVPD_paper_katagiri}

\end{document}